\documentstyle[psfig]{ioplppt}

\newcommand{\beq}{ \begin{equation} }
\newcommand{\eeq}{ \end{equation} }
\newcommand{\bea}{ \begin{eqnarray} }
\newcommand{\eea}{ \end{eqnarray} }
\newcommand{\hypsech}{\mbox{$\rm{sech}(\pi t/ \tau)$}}
\newcommand{\tauovertwopi}{\frac{\tau}{2\pi}}

\begin{document}
\jl{2}
    \title{
     Ionization dynamics in 
     intense pulsed laser radiation. Effects of frequency chirping.
    }
    \author{R Marani\footnote{ Present address : Institut d'Optique, BP 147,
91403 Orsay, France } and E J Robinson}
    \address{Physics Dept., New York University, New York, NY 10003, USA}
    \date{}
\begin{abstract}
Via a non--perturbative method we study the population dynamics and 
photoelectron spectra of
Cs atoms subject to intense chirped laser pulses, with gaussian beams. 
We include above threshold ionization spectral peaks.
The frequency of the laser is near resonance with the 6s-7p transition.
Dominant couplings are included exactly, weaker ones accounted for
perturbatively.
We calculate the relevant transition matrix elements, including
spin--orbit coupling.
The pulse is taken to be a hyperbolic secant in time and the chirping
a hyperbolic tangent. This choice allows the
equations of motions for the probability amplitudes to be solved
analytically as a series
expansion in the variable $u=(\tanh(\pi t/\tau)+1)/2$, where $\tau$ is a
measure of the pulse length.
We find that the chirping changes the ionization dynamics and the
photoelectron spectra noticeably, especially for longer pulses of the
order of $10^4$ a.u.
The peaks shift and change in height, and interference effects between
the 7p levels are enhanced or diminished according to the amount of
chirping and its sign.
The integrated ionization probability is not strongly affected.
\end{abstract}

\section{Introduction}

The ionization of an atom by a pulse of laser radiation is a
fundamental process which is only well understood in certain regions
of intensity, frequency and pulse length. 
The regime which is best understood is when perturbation theory may be
applied, that is where the intensity is very low, the frequency not
too low, and the pulse very long and constant or quasi constant
in electric field amplitude (see e.g.
\cite{delo-krai:84}).
In particular, the amplitude of the initial state is approximately
equal to one for the entire duration of the process.

Since the intensity of the radiation field is nowadays often comparable
with the internal field of the atom, perturbation theory is
inadequate. 
Perturbation theory also 
breaks down at lower intensities if the field frequency is
close to a resonance, or if saturation occurs.
Thus, the applicability of lowest--order theory is limited to
non--resonant absorption. 
Resonance effects involving
intermediate states can make the picture much more complicated and
must be described by a theory which allows for strong signals.

Another degree of complexity appears when the length of the pulse is shortened
(less than $10^{-13}$ s) : the peaks shift, widen and develop substructures.
For example the energies of the atomic states shift by amounts,
which, to the first non--vanishing order, are proportional
to the intensity of the laser field (A.C. Stark effect). 
Thus an intermediate atomic level can shift in such a 
way that it comes into resonance at certain times during the laser pulse.

The above processes have been approached theoretically in a few different ways
and described in different, not always compatible languages, often 
corresponding to non--identical experimental situations.

Some calculations are based on the application of {\it Floquet theory }
\cite{potv-shak:90,dorr-etal:92}, a semiclassical approach which takes
advantage of the temporal periodicity of the Hamiltonian. 

Other theories are based on the {\it Keldysh--Faisal--Reiss} (KFR) 
model \cite{reiss:80,reiss:90}.
In an approximation originally due to Keldysh and developed by Reiss,
the exact final state of the electron is replaced by
solutions of the Schr\"odinger equation for a free electron in a laser
field.
The Keldysh approximation may be  quite successful if the atomic potential is
of very short range, at least for the position of spectral peaks. 
However KFR treatments have not fared particularly well
in quantitative comparisons with experiments specifically designed to test their
predictions regarding ATI electron spectra \cite{peti-etal:88}.

Another type of method is the {\it essential states} approach 
\cite{eber-etal:91}, which focuses
on continuum--continuum interactions.
States are called ``essential'' if they are populated during the entire
process of ATI. 
Such a theory allows simple analytical solutions, provided  
some assumptions are made. These include the rotating wave approximation,
the treatment of transitions into the continuum from the ground state as a direct
non--resonant  multiphoton process, and the assumption that all transition
matrix elements between states belonging to different continua have
one of several simple analytic forms. 

In a number of studies, the time--dependent Schr\"odinger equation has
been solved by {\it direct numerical integration } for model atoms.
Studies have been done both on one--dimensional atoms (e.g.
\cite{java-etal:88,geltman:94}), and more realistic models
\cite{kulander:88}, using various approximations.
One of the limitations of this hard numerical
approach is the time required for these simulations and attendant lack of
flexibility. For example
the space--temporal shape of the laser beam can only be roughly taken
into account. 
This approach does not offer great insight into the physics, but since the
experimental results can vary greatly even for very similar
experiments, these results can be looked at as numerical experiments
against which one can further test model theories. 

The above--mentioned models usually represent only some 
important features of real systems and make simplifying assumptions. 
Many studies are done for
constant fields, or cannot describe processes of resonant--enhanced 
multiphoton ionization.
The same can be said about numerical computations, very sophisticated and 
time consuming even for one--dimensional atoms. 
They only partially contribute to a full understanding of the processes.

Due to the complexities involved in the dynamics of the interaction of
an atom with strong short laser pulses, it is useful to try to 
unveil such dynamics using a model which is both realistic and offers a
simple physical interpretation.

We report on a model of the system which is solvable, at least
partially, by analytical methods, since these generally allow greater 
insight than a direct numerical integration.
This also reduces the computer effort per parameter set, thus
permitting  the integration over spatial profiles, where many
repetitions are needed.

The work is performed in the semiclassical approximation which has given
very good results in the past and is justified given the high
intensity of the radiation field.
Furthermore the electric dipole approximation is indeed almost always valid 
at optical frequencies.

A starting point is the Rosen--Zener problem \cite{rose-zene:32,robiscoe:78,adle-etal:95a}.
In this problem a two--level atom is subject to a pulse whose
amplitude
that varies in time as a hyperbolic secant. Such problem can be
solved
analytically via a transformation of the time variable $t$ to the ``compressed
time'' $u=(\tanh(\pi t/\tau)+1)/2$.
The solutions are hypergeometric functions.

Our model can be regarded as an extension of the basic Rosen--Zener
problem to an atom with more that two active levels, including coupling to the
continuum, to model ATI, and chirping of the laser. 

By using an expansion of the wave function in the amplitudes of atomic states that are
most relevant to the dynamics of the system, one obtains a simplification
that allows one to get a clearer picture of what is going on during the
interaction.
These levels are those most strongly coupled to the ground state, and
others coupled to that group.
In other words, we use a truncated spectral representation of the wave function, which
considers only the atomic states most strongly mixed with the ground state and 
includes the contribution of the remaining levels approximately.

In the application of the method we focus on the cesium atom, both
because it is advantageous theoretically, since many of the atomic
parameters involved have already been calculated in the literature, and
it is also an element which is accessbile to experimental analysis. 

We are interested in studying ATI, thus the coupling of the bound
states to the continua corresponding to the second
ionization peak will also be included.
We also include frequency chirping in the laser pulse and account for
a Gaussian spatial shape.

Section \ref{sec:2} introduces the model for the physical system under
study and the method of solution of the time dependent Schr\"odinger
equation is discussed. 
In section \ref{sec:3} the atomic and laser parameters are characterized.
The results of the simulations are presented
in section \ref{sec:4} and the conclusions in section \ref{sec:5}.
\section{Alkali Atoms in Strong Pulsed Resonant Laser Radiation}
\label{sec:2}

In this section we describe the theoretical model and the solutions
of the corresponding equations of motion. Atomic units are used
throughout this article, that is, one takes $e=m=\hbar=1$ a.u.

\subsection{The Model}

The system under study consists of an alkali atom, which has a
single active electron, interacting with
pulsed laser radiation.
The electron wave function is expanded in the basis of
unperturbed energy eigenstates. To simplify the notation the atomic
levels are labeled with a single index, which stands for the standard set of quantum numbers
completely identifying the electron state. 
In this notation the electron wave function can be written as
\beq
|\Psi(t)>=\sum_n c_n(t) \exp(-i\omega_n t)|n> 
\label{eq:1}
\eeq
where the sum spans the bound states and includes an integration over the 
continuum.

As noted in the introduction, we make the dipole approximation. In addition, the
work is done in the length gauge because it can be argued that it gives
a better approximation than the velocity gauge when a {\em truncated} basis
is used to represent the electronic wave function. 
If the unperturbed potentials are velocity--independent, the momentum matrix
elements are related to the dipole matrix elements by
$<n|p|0> = i(E_n -E_0)<n|z|0>$.
Thus, the momentum matrix elements decrease 
more slowly with the energy difference than do the length ones. 
Then, calculating multiphoton
transition amplitudes, in the sum over intermediate states, the velocity
gauge weighs states more distant in energy more than the length
gauge does and the sum converges more slowly.
Thus, a truncated basis using states close to resonance is a better 
approximation in the length gauge.
The Hamiltonian in length gauge is given by
\beq
H= H_0 + {\bf E}\cdot{\bf r}
\label{eq:2}
\eeq
where $H_0$ is the unperturbed atomic Hamiltonian.

The incident laser field can be written as
\beq
{\cal E}(t) = 2 E(t) \cos(\Omega t + \alpha (t))
\eeq
where $E$ represents the pulse shape and $\alpha$ the frequency chirping,
i.e. a time-dependence of the optical frequency.
Experimentally
this effect usually accompanies the creation of short laser pulses, and
to the best of our knowledge, has not been included in previous
work on ionization.

\begin{figure}
\centerline{\psfig{figure=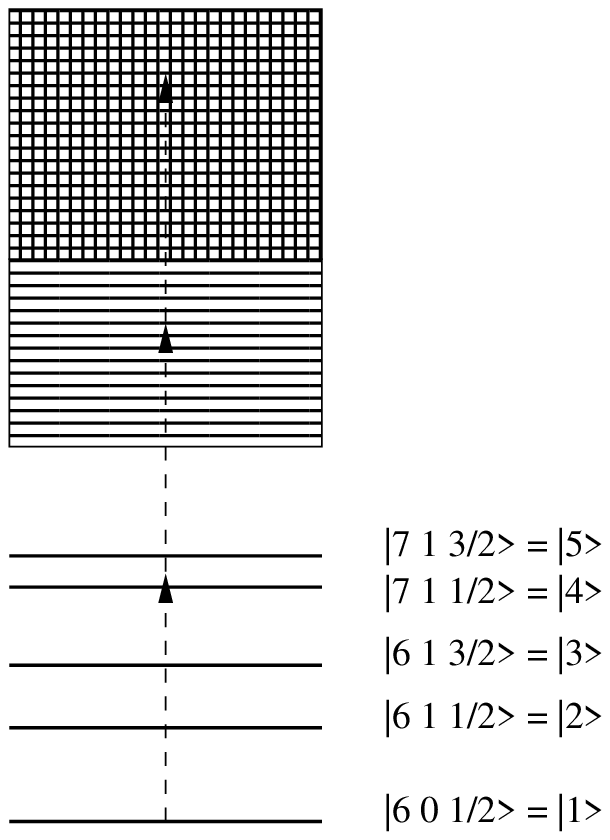}}
\caption{Energy level structure for the Cesium five level system, 
with the excited states coupled to a continuum of states}
\label{fig:2.1}
\end{figure}

We consider photoionization processes in which the laser
frequency is nearly resonant with the transition between the ground
and some excited levels, so that ionization is possible from the
excited states through the absorption of one photon.
Specifically we will study cesium subject to laser light in resonance
with the 6s--7p transition. 
The level scheme is shown in Fig. \ref{fig:2.1}.


If the frequency of the light is close to resonance for the transition
between the ground state and some excited levels, the dynamics of the
electron will be dominated by the corresponding contributions of such levels. In
addition, other levels should be retained if, although nonresonant,
they have a very strong coupling with the ground state. 
We call all these states ``relevant'' states and solve for their effect
exactly.
In addition we can correct for the contribution of the further atomic levels 
in a perturbative way, calculating the energy shifts they induce on
the relevant states.
We use the rotating wave approximation, since we are interested 
in optical frequencies, a regime where such an approximation is
typically excellent.

We make the Weisskopf--Wigner approximation so that the coupling
between the bound states and the continuum levels is modeled by 
decay coefficients and level shifts. This can be done also for the
processes corresponding to excess photon absorption (i.e. ATI).

We now give an outline of the derivation of the equations of motion for the 
probability amplitude using the afore mentioned approximations.
Using Eqs.(\ref{eq:1}) and (\ref{eq:2}) and assuming the electric field to be 
polarized in the $\hat z$ direction, 
the time--dependent Schr\"odinger equation yields:
\beq
i \dot c_m = {\cal E}(t) \sum_n \kappa_{mn}\exp(-i\omega_{nm}t) c_n(t)
\eeq
where $\omega_{mn}\equiv\omega_m-\omega_n$ and 
$\kappa_{mn}\equiv<m|z|n>$.
Initially the atom is its ground state. 
Since we assume that the non--relevant states and the continuum are going
to be scarcely populated, we can treat them perturbatively, so that
for one of such states, $k$:
\beq
c_k(t)\approx -i\sum_n \kappa_{kn} \int_{-\infty}^t dt' {\cal E}(t')
\exp(-i\omega_{n k}t')c_{n}(t'),
\eeq
The sum is over the relevant states.
We then substitute the above expression into the equation of motion
for the relevant state $m$ and perform the time integration by parts,
we get a term containing a time derivative which we can neglect in
the Weisskopf--Wigner approximation.

Further we neglect terms containing factors such as $\exp(2i\Omega t)$.
We then assume that $\Omega\approx\omega_{nm}$ (quasi--resonant 
approximation), keep only the resonant term explicitly and the
first non--vanishing off--resonance correction. 
Finally, we set $c_m = a_m\exp(-i\lambda_m\int_{-\infty}^t dt' E^2(t'))$.

With these approximations the equations of motion for the probability
amplitudes $\{ a_i, i=1,N \}$ are 
\begin{eqnarray}
\frac{d a_1}{dt} &=& -i\sum_{l=2}^{N} \kappa_{1l} E(t) 
 \exp\left[i(\Delta_{1l} t + \phi_{1l}(t)+\alpha(t))\right] a_l(t),  \\
\frac{d a_m}{dt} &=& -i\kappa_{1m} E(t) 
 \exp\left[-i(\Delta_{1m} t +  \phi_{1l}(t)+\alpha(t))\right] a_1(t)
 \\\nonumber
    && - [\gamma_m E(t)^2/2] a_m(t), \\
\phi_{lj}(t)&\equiv&(\lambda_l-\lambda_j)\int_{-\infty}^{t} dt' E(t')^2
\end{eqnarray}
where $\kappa_{ij}$ are the coupling coefficients between levels $i$ and
$j$, i.e. $\kappa_{ij}=<i|z|j>$,
$\Delta_{ij}\equiv \Omega - (\omega_j -\omega_i)$ is the detuning from the
transition frequency, the $l$ summation is
over all excited states, $\lambda_j$ is the Stark coefficient for the $j$th
level and $\gamma_j E(t)^2$ is the decay rate, i.e.
$\gamma_j = 2\pi |\kappa_{jk}|^2$, with $\omega_j-\omega_k=\Omega$.
Since we will consider only a small range of frequencies near
resonance, we assume that the decay coefficients are constant over
this energy interval.
The explicit Stark shift terms include contributions from all other levels and also
the non--resonant contribution from the same level, which provides a
correction to the rotating wave approximation. For an excited level $j$,
the Stark shift is given by
\beq
\lambda_j=\frac{|z_{j1}|^2}{\Omega + \omega_{j1}} +
{\sum_k}' \left(\frac{|z_{jk}|^2}{\Omega + \omega_{jk}} +
\frac{|z_{jk}|^2}{\omega_{jk}-\Omega} \right),\: (j\neq 1)
\label{eq:starkj}
\eeq
and by
\beq
\lambda_1=\sum_n\frac{|z_{1n}|^2}{\Omega + \omega_{1n}} +
{\sum_k}' \left(\frac{|z_{1k}|^2}{\Omega + \omega_{1k}} +
\frac{|z_{1k}|^2}{\omega_{1k}-\Omega} \right)
\label{eq:stark1}
\eeq
for the ground state,
where the sum over $k$ includes the principal value of the integral
over the continuum states as well as the bound states not explicitly
considered in the model, while the sum over n spans the ``relevant'' states.
The equations for the continuum amplitude for a state of frequency
$\omega_{\kappa}$ are 
\begin{equation}
\frac{d a_{k}}{dt} = -i E(t) \sum_{l=2}^{N} \kappa_{lk}
 \exp\left[-i\left(\Delta_{lk} t + \phi_{lk}(t) + \alpha(t)\right)
     \right] a_{l}(t)
\end{equation}
for the first ionization peak, and
\begin{equation}
\frac{d a_{q}}{dt} = -i E(t)^2 \sum_{l=2}^{N} \chi_{lq}
 \exp\left[-i\left(\Delta'_{lq} t + \phi_{lq}(t) + \alpha(t)\right)
  \right] a_{l}(t)
\end{equation}
for the second ionization peak (ATI). Here the detuning is given by
$\Delta'_{lq}=2\Omega-(\omega_q-\omega_l)$, the Stark coefficient
$\lambda_k\equiv 1/\Omega^2$ and $\chi_{lq}$ is the two--photon
coupling between the bound state $l$ and the continuum state $q$ (see
Eq.(\ref{eq:chi2}).

In the above equations we consider only the lowest order terms in the
field, i.e. cross terms corresponding to
the absorption of one photon from an excited state and re--emission to
a different one are neglected. A few test runs have shown this contribution
to be negligible. In the same manner, we neglect the effect of the 
decay rate to the
second continuum on the equations of motion for the
excited bound states. One could readily allow for this effect, but it
proves to be negligible for the conditions we encounter. We also neglect the process involving
absorption of two photons from an excited state and re--emission of one
photon to a continuum state. 

While the equations can be generalized to any number of excess photons
absorbed, we will consider only the first two peaks.

To enable the analysis to be carried out in closed form, we chose a field envelope in the shape of a
hyperbolic secant and a frequency chirp proportional to a hyperbolic
tangent :
\beq
E(t) = \sqrt{I_0}\hypsech, \; 
\alpha(t) =\beta\int_{-\infty}^t dt' \tanh(\pi t'/\tau)
\eeq
where the characteristic time, $\tau$, of the amplitude modulation is
large compared to $1/\Omega$.
This allows a solution in the form of a power series in
the compressed time, which we will define below.


\subsection{Solution of the Equations of Motion}

In order to solve the equations of motion for the probability amplitude
given in the previous section, we first let
\bea
b_1&=&a_1\\
b_j&=&\exp[i(\Delta_{1m} t + \phi_{1j}(t)+\alpha(t))]a_m,
\eea
where $m=2,3,\dots,N$.
The equations of motion for the new $b$ amplitude are
\begin{eqnarray}
\frac{d b_1}{dt} &=& -i\sum_{l=2}^{N} \kappa_{1l} E(t) b_l(t),  \\
\frac{d b_m}{dt} &=& -i\kappa_{1m} E(t) b_1(t) - \frac{\gamma_m E(t)^2}{2} b_m(t)
\nonumber \\
 && +i\left(\Delta_{1m} + \dot\phi_{1m}(t)+\dot\alpha(t)\right)b_m,
\end{eqnarray}
where
\bea
\dot\phi_{lm}&=&(\lambda_l-\lambda_m)E(t)^2,\\
\dot\alpha(t)&=&\beta\tanh(\pi t/\tau).
\eea
The continuum probability amplitudes are found by formally integrating
the corresponding equations of motion:
\beq
a_{k}(t) = -i \int_{-\infty}^t dt' E(t') \sum_{l=2}^{N} \kappa_{lk}
 \exp\left[-i\left(\Delta_{1k} t' + \phi_{1k}(t')+\alpha(t')\right)
\right] b_{l}(t')
\end{equation}
for the first ionization peak, and
\begin{equation}
a_{q}(t) = -i \int_{-\infty}^t dt' E(t')^2 \sum_{l=2}^{N} \chi_{lq}
 \exp\left[-i\left(\Delta'_{1q} t' + \phi_{1k}(t')+\alpha(t')\right)
\right] b_{l}(t')
\end{equation}
for the second ionization peak (ATI). 
 
Letting $u\equiv(\tanh(\pi t/\tau)+1)/2$ and expanding the bound state amplitudes
in a power series in $u$, we find :
\beq
b_1=\sum_{n=0}^\infty \alpha_n u^n, \;  b_j=\sum_{n=0}^\infty \beta_n^{(j)}
u^n\hypsech\equiv\sum_{n=0}^{\infty}\mu_n^{(j)}u^n.
\eeq
The expansion coefficients can be calculated through the following
recursion equations:
\beq
(j+1)\alpha_{j+1} + i \frac{2\tau}{\pi} E_0 \sum_{l=2}^N 
\kappa_{1l}\beta_j^{(l)}=0
\eeq
\begin{eqnarray}
&&\left[n+{1\over 2} -i\tauovertwopi(\Delta_{1j}-\beta)\right]\beta_n^{(j)}=
\left[n+i\tauovertwopi
(4\lambda_{1j}I_0 +2\beta) - 
\right. \nonumber \\
&& 
\left.2 \gamma_j E_0^2
\tauovertwopi\right]\beta_{n-1}^{(j)} 
+(-i \lambda_{1j} I_0 4\tauovertwopi + 2\gamma_j I_0
\tauovertwopi)\beta_{n-2}^{(j)}
-i\tauovertwopi\kappa_{1j}\sqrt{I_0}\alpha_n.
\end{eqnarray}

Using the expansion in powers of $u$ in the formula for
the continuum amplitude and integrating term by term, we get
\bea
a_{k} &=& -i \frac{E_0\tau}{\pi} 2^{i\beta\tau/\pi} \sum_{n=0}^\infty
\sum_{l=2}^N \mu_n^{(l)} \kappa_{lk}  \nonumber \\
&\times& B\left(\frac{1}{2}+i\tauovertwopi(\Delta_{1k}+\beta),
n+1-i\tauovertwopi(\Delta_{1k}+\beta)\right) \nonumber \\
&\times& _1\!F_1\left(n+1-i\tauovertwopi(\Delta_{1k}-\beta),\frac{3}{2}+n
+i\tauovertwopi 2\beta,-i(\lambda_1-\lambda_k)I_0\frac{2\tau}{\pi}\right)
\eea
and
\bea
a_{q} &=& -i \frac{I_0\tau}{\pi} 4^{i2\beta\tau/\pi} \sum_0^\infty
\sum_{l=2}^N \mu_n^{(l)} \chi_{lq} \nonumber \\
&\times& B\left(1+i\tauovertwopi(\Delta'_{1q}+2\beta),
n+\frac{3}{2}-i\tauovertwopi(\Delta_{1k}-2\beta)\right)  \nonumber \\
&\times& _1\!F_1\left(n+\frac{3}{2}-i\tauovertwopi(\Delta_{1k}-\beta),\frac{5}{2}+n
+i\tauovertwopi 4\beta,-i(\lambda_1-\lambda_k)I_0\frac{2\tau}{\pi}\right).
\eea
where $B$ is the beta function and $_1F_1$ the confluent hypergeometric function.
These results have been implemented both with {\em Mathematica} and
Fortran programs. {\em Mathematica} allows an arbitrary amount
of precision, but it is about a hundred times slower than the Fortran
version. For fields up to $10^{-5}$ atomic units and pulse lengths up to
$10^4$ atomic units Fortran's double precision proved to yield the
same results as the {\em Mathematica} code.

\section{Atomic and Laser Parameters}
\label{sec:3}

The model atom is characterized by several
coupling and decay constants and level shifts,
which need to be evaluated. 
Coefficients for one--photon transitions are given
by the matrix elements $\kappa_{fi}=<f|z|i>$ and the Stark--shift
parameters $\lambda_j$ by Eq.(\ref{eq:starkj}) and (\ref{eq:stark1}). The two--photon
coupling between the excited states and the continuum is given by
$\chi_{fi}$ in Eq.(\ref{eq:chi2}). 
For the case of the cesium atom the above parameters are already available 
in the literature (see e.g. \cite{adle-etal:95a})
except for the two--photon couplings above threshold.

We choose the laser field to be linearly polarized in the z
direction and of frequency such that we can model this problem 
accurately with the retention of only four excited states, those of
the 6p and the 7p doublets.
The 7p states are quasi resonant, while the 6s-6p oscillator strengths are
orders of magnitude larger than any other coupling to the ground state.
Thus, retaining the 6p doublet takes non resonant ionization into account 
with negligible error. The level scheme is shown in Fig. \ref{fig:2.1}.
We take the ground state energy to be equal to zero.
The ionization potential of cesium is 3.89 eV.

In Table \ref{tab:3.1}, we list the energy levels and Stark shift
coefficients of the relevant states of cesium.

\begin{table}
\caption{ Energy levels and Stark shift coefficients of the relevant
states of cesium}
\label{tab:3.1}
\lineup
\begin{indented}
\item[]\begin{tabular}{@{}lll}
\br
State   & Energy (a.u.) & Stark Shifts (a.u.)\\
\mr
6$^2$s$_{1/2}$ & $\omega_1$=0.0      & $\lambda_1$= -76.93493\\
6$^2$p$_{1/2}$ & $\omega_2$=0.050931 & $\lambda_2$= 159.221\\
6$^2$p$_{3/2}$ & $\omega_3$=0.053456 & $\lambda_3$= 232.48688\\
7$^2$p$_{1/2}$ & $\omega_4$=0.0992   & $\lambda_4$= 64.3317\\
7$^2$p$_{3/2}$ & $\omega_5$=0.1000   & $\lambda_5$= 386.861\\
\br
\end{tabular}
\end{indented}
\end{table}

The decay parameters were calculated in \cite{adle-etal:95a} by
the method developed by Burgess and Seaton \cite{burg-seat:60},
the matrix elements coupling of the bound states were calculated
using the quantum defect method and the non resonant couplings were
calculated using data found in Stone \cite{stone:62}.
In the present work we also need the coupling
coefficients between the bound states and the continua associated
with the second ionization.
This corresponds to the absorption of 2 photons from the excited
states, when the absorption of 1 photon is permitted (that is, we have
``above threshold ionization'').

\subsection{Calculation of the coupling to ATI channels}

One needs to evaluate terms like :
\beq
\chi_{fi}=\lim_{\eta\to 0}\sum_k
\frac{z_{fk}z_{ki}}{\Omega+\omega_i-\omega_k +i\eta}.
\label{eq:chi2}
\eeq
The sum spans all intermediate bound and continuum states. Here
$\Omega$ is the laser frequency and $\omega_i$ is the energy of the
initial bound state.
The $\eta$ in the denominator was included because above threshold
($\Omega>|\omega_i|$)
the denominator vanishes when $\omega_n = \omega_i +\Omega$.
The small imaginary part prescribes the correct boundary conditions
for a field adiabatically turned on in the remote past.

Several methods have been developed to calculate transition amplitudes
involving infinite sums over intermediate states.
For alkali atoms, quantum defect theory and model potentials have
been used as weel as truncated summations over a finite number of states.
Most calculations have taken the laser frequency to be below
threshold.

We used a method based on the solution of an Inhomogeneous
Differential Equation (IDE), which is equivalent to performing the sums over
the entire spectrum of bound and continuum intermediate states.
In this formulation we have 
\begin{equation}
\chi_{fi} = <\Psi_f|z|\tilde \Psi_i>
\end{equation}
where $|\tilde\Psi_i>$ is the solution of
\begin{equation}
[ H_0- \omega ]\tilde\Psi_i(\mbox{\bf r}) =
- z \Psi_i(\mbox{\bf r}),
\label{IDE}
\end{equation}
with $\omega= \omega_i +\Omega$, which is a positive quantity for ATI.

This method, which is based on numerical computations in the framework
of the central field approximation has a wider range of applicability
than the afore mentioned analytical methods, and also leads to
accurate results. 

The central potential we used is calculated in \cite{herm-skil:63} using the
Hartree--Fock--Slater approximation. We add a spin--orbit interaction of
the form 
\beq
V_{SO}(r)= {\bf \hat L\cdot \hat S}\frac{\alpha^2}{2r}\frac{dV}{dr}
\eeq
where $V$ is the Hartree--Fock potential calculated without
spin--orbit interaction.
We work in the SLJM representation.

The initial state wave function $\Psi_i$, represents a state identified
by the quantum numbers $i\equiv\{n_i,j_i,l_i,m_i\}$ and can be expressed in terms
of the product of a radial function and a generalized spherical harmonic :
\begin{equation}
\Psi_i(r,\hat\Omega) = R_{i}(r) {\cal Y}(j_i,l_i,m_i;\hat\Omega).
\end{equation}

We look for a solution in terms of a superposition of eigenstates of
the total and orbital angular momentum.

\begin{equation}
\tilde\Psi_i(\omega,r,\hat\Omega)= 
\sum_{j,l,m} \tilde R_i(\omega,j,l,m;r){\cal Y}(j,l,m;\hat\Omega).
\end{equation}
Projecting Eq.(~\ref{IDE}) on state $|jlm>\equiv {\cal Y}(j,l,m;\hat\Omega$), 
we obtain :
\begin{equation}
[ H_0- \omega ]\tilde R_i(\omega jlm;r)=-<jlm|\cos\theta|j_i l_i m_i> 
R_{n_i}(j_i l_i m_i;r)r.
\end{equation}

The matrix element of $\cos \theta$ can be calculated
by an application of the Wigner--Eckart theorem \cite{lamb-teag:76} :
\bea
<l'j'm'|r^q|ljm>&=&(-1)^{3/2+j'+j-m'-l_>} \sqrt{l_>} 
[(2j'+1)(2j+1)]^{1/2}r \nonumber\\
&&\times
\left( \begin{array}{ccc}
j'& 1 & j \\
-m' & q & m 
\end{array}\right)
\left\{\begin{array}{ccc}
l' & j' & 1/2 \\
 j & l & 1
\end{array}\right\}
\eea
where $r^0=z, r^{\pm 1} = \mp 1/2 (x \pm i y)$ and $l_> \equiv \max(l,l')$.
In our case $q=0$. 

For cesium the ground state is 6S$_{1/2}$.
Since the final result will not depend on the projection of the spin, we can 
pick a state with m$_S$=1/2 without loss of generality, i.e. 
n= 6, l=0, j=1/2, m=1/2. Since the interaction $z$ conserves the spin and
m$_l$, m will always be equal to $+1/2$ and we will drop it in subsequent
formulas.

After the absorption of one photon from the ground state, 
the electron will be in a state p$_{1/2}$
or p$_{3/2}$. Since we will consider laser frequencies close to the
transition frequency between the ground state and the 7p's states, our
atomic model will consider these states explicitly, and as noted, also 
the 6p's, because their coupling to the ground state is very large.
Thus, the initial state for the transition to the continuum via the
absorption of 2 photons can be any of the 6p's or 7p's states.
Let $P(r) \equiv r R(r)$, then
\bea
&&\left[ - \frac{d^2}{dr^2} +\frac{l(l+1)}{r^2}+V(jl;r)-\omega\right]
\tilde P_i(\omega j l, r)= \nonumber\\ 
&&-<j l {1\over 2}|\cos\theta|j_i l_i {1\over 2}>r P_n(j_i,l_i;r).
\label{eq:IDE}
\eea

where $R_n$ is an eigenstate of the unperturbed atom with $n=6,7$,
$j=1/2,3/2$ and $l_i=1$.

First we solve the unperturbed Schr\"odinger equation to calculate the
initial bound state which enters in the inhomogeneous term of
Eq.(\ref{eq:IDE}). This is done through a modification of Herman and
Skillman's code ~\cite{herm-skil:63} to include spin orbit
interaction.

We solve Eq.(\ref{eq:IDE}) numerically using Numerov's method 
(see e.g. \cite{koon-mere:b90}) integrating outward from the origin.
The boundary conditions require the solution to be zero at the origin
and to be an outgoing wave for large $r$.
We start the integration using a Taylor expansion around the origin.
In general the solution obtained does not satisfy the boundary
conditions at infinity, so we add a solution of the homogeneous
equation such that the combination satisfies the boundary conditions
at infinity. The coefficients of the combination are determined using
the asymptotic expansion of Coulomb functions. In fact, at a large
distance from the origin, the potential approaches $-1/r$ so that
the solutions have the form of phase--shifted Coulomb waves.

The transition matrix to a final state $|\omega_q,l_f,j_f>$
is given by
\begin{eqnarray}
&&<\omega_f,l_f,j_f | z | \tilde \Psi_i({\omega})>=
\sum_{l'j'}
<\omega_f,l_f,j_f | z |l'j'><l'j'| \tilde
\Psi_i(\omega)>\\ 
&&=\sum_{l'j'} \int dr r P_{\omega_f}(j_f,l_f;r)\tilde P_i(\omega j'l';r)
<j_fl_f|\cos(\theta)|j l> .\nonumber
\end{eqnarray}

The radial integral is calculated by dividing the integration range in two
parts. The integration is done numerically up to a certain distance
$R_{\max}$. The remaining integral to infinity is evaluated using 
an expansion in terms of inverse powers of $r,n$ and, 
whose asymptotic expansion is given in \cite{abra-steg:94} as
\bea
F_l(r)&=&f(r)\cos\theta_l - g(r)\sin\theta_l,\\
G_l(r)&=&f(r)\sin\theta_l + g(r)\cos\theta_l,\\
\theta_l &=& kr+\ln(2kr)/k-l\pi/2+\sigma_l,\\
\sigma_l&=&\arg\Gamma(l+1-i/k),
\eea
where $f$ and $g$ can be expanded in inverse powers of $kr$ for large
$kr$ ($k=\sqrt{2E}$). The resulting integrals have an
analytical form, which maybe written as 
\bea
\int_R^{\infty} &&dx \exp(i a x) x^{\lambda -i b} = \nonumber \\
&&- \frac{\exp(i a R)}{i a} R^{\lambda -i b} \left( 1 -
\frac{\lambda -ib}{i a R}+\frac{(\lambda -i b)(\lambda -i
b-1)}{(iaR)^2} - ... \right).
\eea 

\subsection{Transition Amplitudes for Cesium and Laser Parameters}

\begin{figure}
\centerline{\psfig{figure=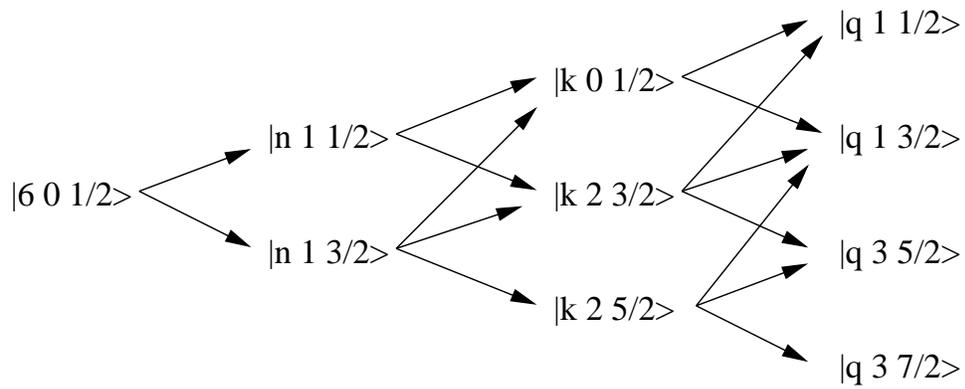,height=2in}}
\caption{Ionization channels from the ground state of cesium}
\label{fig:channels}
\end{figure}

By virtue of the angular momentum selection rules, the 6s ground state
couples to the 6p and 7p states, while the p states couple via one--photon
absorption to the s and d continuum channels and to the p and f continuum
channels via two--photon absorption.
Fig.~\ref{fig:channels} presents a diagram of the various ionization
channels considered in this study and
Tabs.~\ref{tab:3.2}, \ref{tab:3.3} and \ref{tab:3.4} give the corresponding
coefficients calculated with the methods described in the previous section.

\begin{table}
\caption{ Coupling coefficients between the ground state and the excited
states.}
\label{tab:3.2}
\lineup
\begin{indented}
\item[]\begin{tabular}{@{}ll}
\br
Parameter & Value (a.u.) \\ 
\mr
$\kappa_{12}$ & -1.81 \\
$\kappa_{13}$ & -2.56 \\
$\kappa_{14}$ & 0.11 \\
$\kappa_{15}$ & 0.23 \\
\br
\end{tabular}
\end{indented}
\end{table}

\begin{table}
\caption{ Decay rates from the excited states to the continuum
channels via one--photon transitions, for a laser frequency quasi resonant
with the 6s-7p transition. The corresponding coupling
coefficients are given by $\sqrt{\gamma_{ij}/2\pi}$. }
\label{tab:3.3}
\lineup
\begin{indented}
\item[]\begin{tabular}{@{}lll}
\br 
Parameter & Value (a.u.) & \\ 
\mr
$\gamma_{21}$ & 117.5878 & (6P$_{1/2} \rightarrow {\rm S}_{1/2}$) \\
$\gamma_{22}$ & 10.45479 & (6P$_{1/2} \rightarrow {\rm D}_{3/2}$) \\
$\gamma_{31}$ & 13.83078 & (6P$_{3/2} \rightarrow {\rm S}_{1/2}$) \\
$\gamma_{32}$ & 10.41901 & (6P$_{3/2} \rightarrow {\rm D}_{3/2}$) \\
$\gamma_{33}$ & 93.11069 & (6P$_{3/2} \rightarrow {\rm D}_{3/2}$) \\
$\gamma_{41}$ & 37.4357  & (7P$_{1/2} \rightarrow {\rm S}_{1/2}$) \\
$\gamma_{42}$ & 3.50958  & (7P$_{1/2} \rightarrow {\rm D}_{3/2}$) \\
$\gamma_{51}$ & 3.33084  & (7P$_{3/2} \rightarrow {\rm S}_{1/2}$) \\
$\gamma_{52}$ & 3.003428 & (7P$_{3/2} \rightarrow {\rm D}_{3/2}$) \\
$\gamma_{53}$ & 26.8404  & (7P$_{3/2} \rightarrow {\rm D}_{5/2}$) \\
\br
\end{tabular}
\end{indented}
\end{table}

\begin{table}
\caption{Coupling parameters to the continuum via two--photon
transitions (ATI peak); $i$ is the imaginary unit.}
\label{tab:3.4}
\lineup
\begin{indented}
\item[]\begin{tabular}{@{}lll}
\br
Parameter & Value (a.u.)& \\
\mr
$\chi_{24}$ & $1972-2916i$    & (6P$_{1/2} \rightarrow {\rm P}_{1/2}$) \\
$\chi_{25}$ & $-960+484i$     & (6P$_{1/2} \rightarrow {\rm P}_{3/2}$) \\
$\chi_{26}$ & $-2568+4759i$   & (6P$_{1/2} \rightarrow {\rm F}_{5/2}$) \\
$\chi_{34}$ & $-616-884i$     & (6P$_{3/2} \rightarrow {\rm P}_{1/2}$) \\
$\chi_{35}$ & $3188+2068i$    & (6P$_{3/2} \rightarrow {\rm P}_{3/2}$) \\
$\chi_{36}$ & $844+456i$      & (6P$_{3/2} \rightarrow {\rm F}_{5/2}$) \\
$\chi_{37}$ & $-2672-1356i$   & (6P$_{3/2} \rightarrow {\rm F}_{7/2}$) \\
$\chi_{44}$ & $106.8+29i$     & (7P$_{1/2} \rightarrow {\rm P}_{1/2}$) \\
$\chi_{45}$ & $ 64-15.9i $    & (7P$_{1/2} \rightarrow {\rm P}_{3/2}$) \\
$\chi_{46}$ & $ 383.2-460i$   & (7P$_{1/2} \rightarrow {\rm F}_{5/2}$) \\
$\chi_{54}$ & $62.8-26.84i$   & (7P$_{3/2} \rightarrow {\rm P}_{1/2}$) \\
$\chi_{55}$ & $-142.8+23.92i$ & (7P$_{3/2} \rightarrow {\rm P}_{3/2}$) \\
$\chi_{56}$ & $-65.6+79.6 i$  & (7P$_{3/2} \rightarrow {\rm F}_{5/2}$) \\
$\chi_{57}$ & $ 199.2-235.6i$ & (7P$_{3/2} \rightarrow {\rm F}_{7/2}$) \\
\br
\end{tabular}
\end{indented}
\end{table}

In this work we are interested in short laser pulses.
After the first successful generation of short optical pulses via
mode locking with Nd:glass in the sixties, steady progress in
generating shorter and shorter pulses was achieved.
For a review, see for example \cite{duling:95}.

Nowadays the shortest pulses generated directly from a laser can be
achieved with Ti:sapphire lasers. These lasers can be used to achieve
$10^{-14}$ s pulse generation 
\cite{asak-etal:93} by using nonlinear effects which induce frequency
chirps.
Complete compensation of the chirp is not possible with a
finite number of linear optical elements.

With these experimental considerations in mind,
we will consider laser pulses of durations in the 25-250 fs range, with
frequency almost on resonance with the 6s-7p transition.
The frequency chirping induced by the compression mechanisms, is characterized
by a chirp parameter $\beta$ of the order of $\pi/\tau$.
The parameters characterizing the laser beam are thus frequency, chirp
parameter and intensity. The temporal dependence of the pulse is given
by Eqs.(2.3) and (2.16). We will consider intensities in the range
$10^{-7}-10^{-4}$a.u., and pulse lengths of $10^3-10^4$a.u., with
chirp parameters $\beta\approx \pi/\tau$.


\section{Ionization Dynamics of Cesium}
\label{sec:4}

In this section, the method described in section \ref{sec:2}
 is applied to the case
of cesium, using the atomic and laser parameters calculated and listed
in section \ref{sec:3}. We will consider three different regimes according to the
intensity of the laser beam and the case of Gaussian beams. 
The `atomic' unit of intensity here is
$1.4\times 10^{17}$ W/cm$^2$ -- four times the unit used by many other authors.

\subsection{Weak Field Limit}

The spectrum for weak fields (in the adiabatic limit)
is usually a single peak at an energy
twice the photon energy above the ground state.
In fact, for this case, one expects the final energy of the
photoelectron to consist of the Fourier distribution of the pulse squared
at an energy consistent with energy conservation.
However, with relatively short, weak pulses
tuned between the doublet, a double peak 
structure is found in the s$_{1/2}$ or d$_{3/2}$ continuum channels
but not in the d$_{5/2}$ channel.
This structure was first reported in \cite{adle-etal:95a}
and attributed to the interference between the two different ionization
paths to the afore mentioned channels, which results in the
suppression of the central peak. 
However the d$_{5/2}$ channels can
be reached only in one way, through the p$_{3/2}$ excited states and
thus cannot show any interference effects.

\begin{figure}
\centerline{\psfig{figure=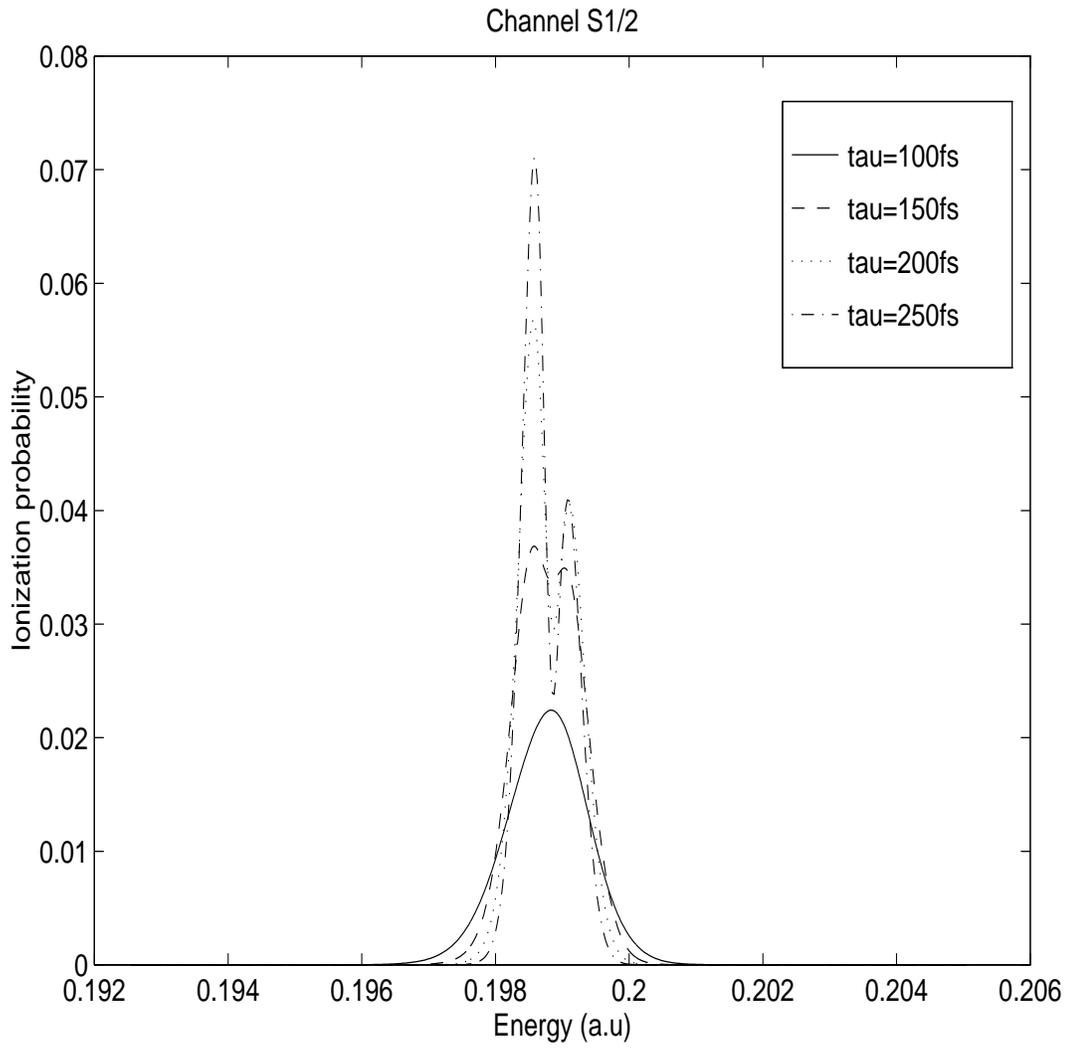,height=5.5in,width=6in}}
\caption{Probability per unit energy vs. energy for the photoelectron
produced by a two photon ionization of cesium 
for various pulse lengths. Laser frequency 0.0994 a.u., peak
intensity $10^{-7}$a.u. ($1.4\times 10^{10}\rm{W/cm}^2$), no chirping.}
\label{sp1_tau1}
\end{figure}

Adler {\em et al.}~\cite{adle-etal:95a} found that one sees a pair of peaks
only when the condition $1<|\Delta_{1j}\tau|<10$ is roughly true.
This behavior is consistent with the interpretation of
such structure as a manifestation of destructive interference 
between the 7p levels, when the laser is tuned to the middle of the
doublet. 
For very short pulses, the incident radiation's spectrum covers
a wider range of frequencies and the detuning is less significant
so that the 7p's are practically a new single state and the doublet
disappears (see Fig.~\ref{sp1_tau1}).
For longer pulses, the central peak forms again and hides the
side peaks, which are much smaller.
The central peak is about at frequency $\omega_1+2\Omega$, while the
two side peaks at $\omega_4+\Omega$ and $\omega_5+\Omega$. The
relative heights are correlated with the average populations of the 7P
levels during the pulse.

\begin{figure}
\centerline{\psfig{figure=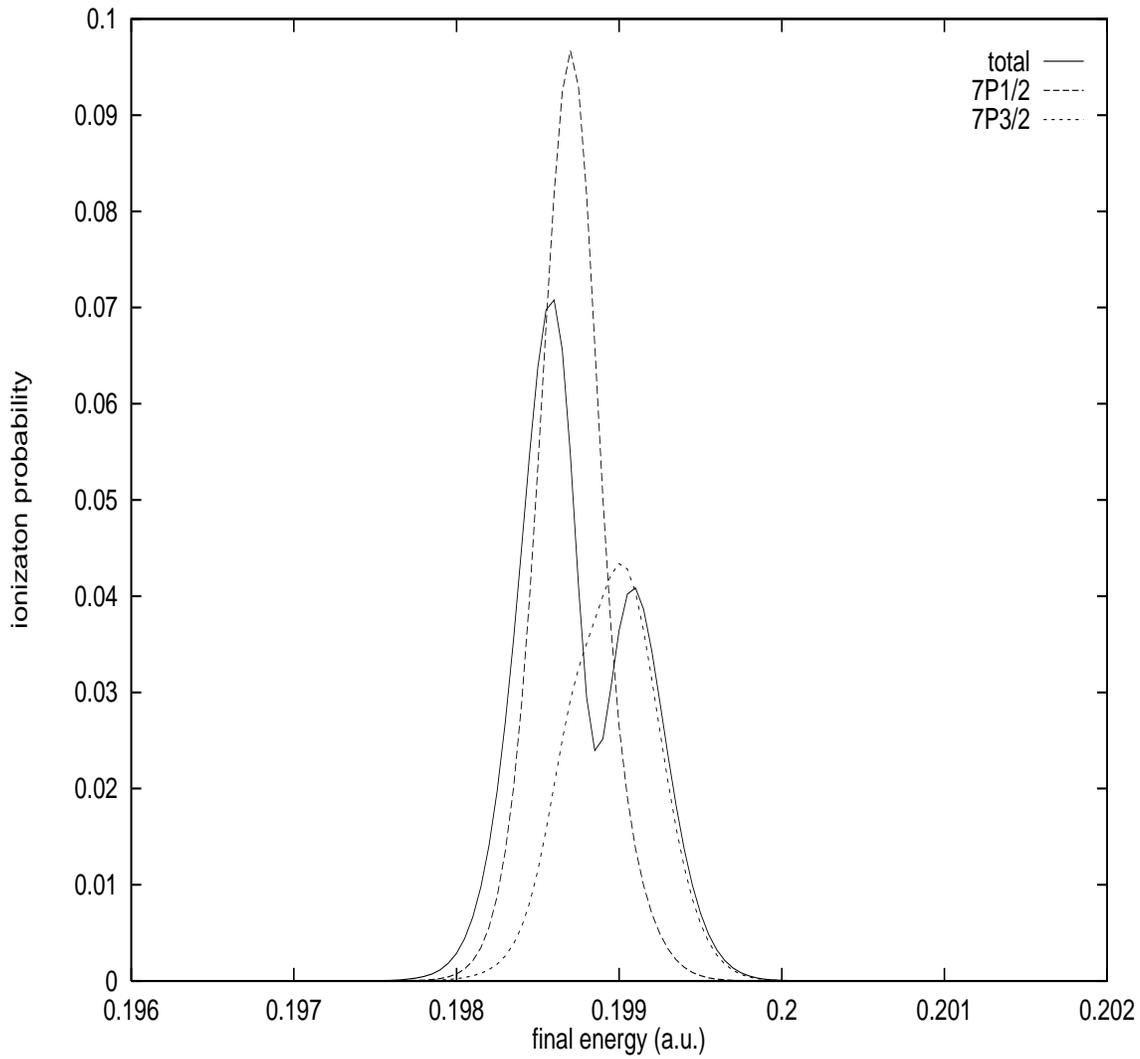,height=5.5in,width=6in}}
\caption{Probability per unit energy vs. energy for the photoelectron
produced by a two photon ionization of cesium. The solid line is the
spectrum for the S$_{1/2}$ channel, the other curves represent the
contribution of ionization from the 7p levels separately (no
interference effects). Laser frequency 0.0994 a.u., peak
intensity $10^{-7}$a.u. ($1.4\times 10^{10}\rm{W/cm}^2$), pulse length 250fs, no chirping.}
\label{sp_separ}
\end{figure}

\begin{figure}
\centerline{\psfig{figure=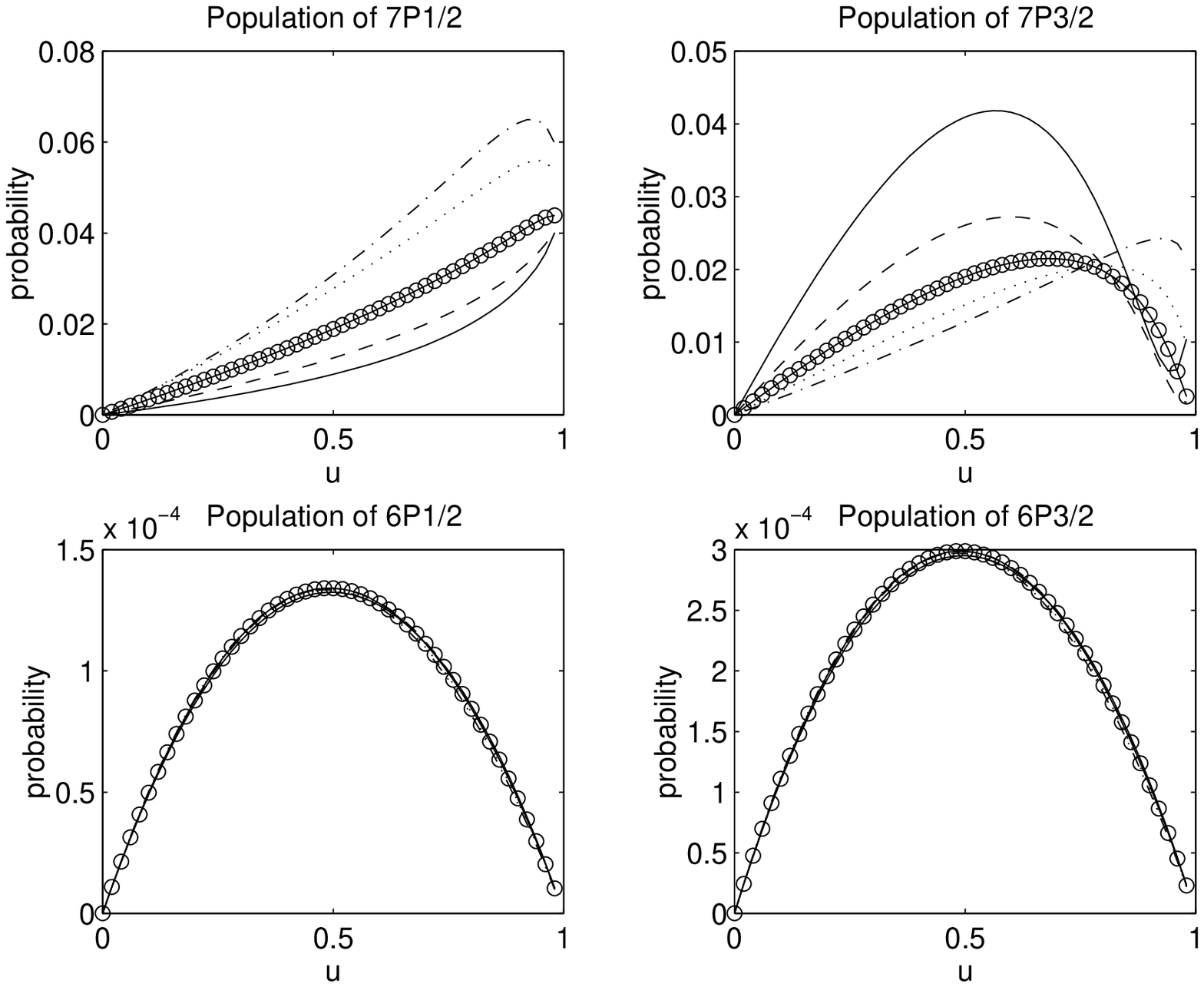,height=6in,width=5.7in}}
\caption{Populations of the 7p levels vs compressed time for
different chirp parameters.
Laser frequency 0.0994
a.u., pulse length 250 fs, peak intensity $10^{-7}$ a.u. ($1.4\times 10^{10}\rm{W/cm}^2$),
line type correspond to different chirp parameters $\beta$:
$-$ -0.0003 a.u., $--$ -0.0002 a.u., o  0.0 a.u.,
$\cdot\cdot$ 0.0002 a.u., $-\cdot$ 0.0003 a.u.}
\label{popP_chirp2}
\end{figure}

In Fig. \ref{sp_separ} we show the spectra of the S$_{1/2}$ channel
and the the spectra corresponding to the contribution from the 7p states
separately, i.e. with no inteference. One can see here that each 7p
state contributes to one of the side peaks. The central peak is not
resolved, being very close to the side peaks. If there were no
interference effects we would not have the dip between the two side peaks.
The existence of two peaks is not due do oscillations in the
populations, since we find that these do not
occur for the intensities we are considering here, as can be seen for
example in Fig.~\ref{popP_chirp2}.

\begin{figure}
\centerline{\psfig{figure=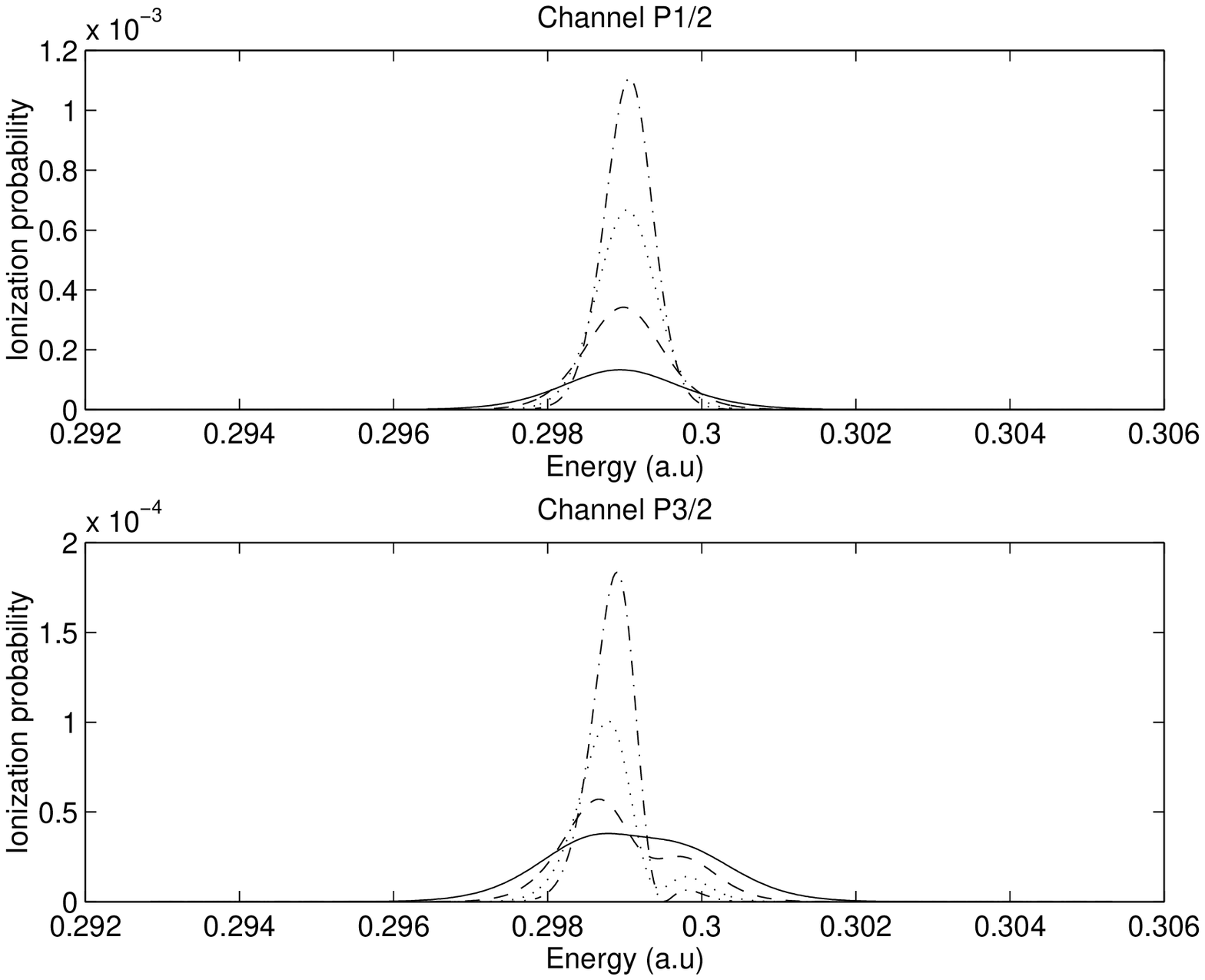,height=6in,width=5.7in}}
\caption{Probability per unit energy vs. energy for the photoelectron
produced by a two photon ionization of cesium 
for various pulse lengths. Laser frequency 0.0997 a.u., peak
intensity $10^{-7}$ a.u. ($1.4\times 10^{10}\rm{W/cm}^2$), no chirping; pulse length $\tau$ is
$-$ 100 fs,$--$ 150 fs, $\cdot\cdot$ 200 fs,$\cdot -$ 250 fs.}
\label{sp3_tau2}
\end{figure}

In the ATI peak the double peak structure is usually absent, in fact
the mediation of several channels to the first continuum, makes
interference effects less likely to occur. For some particular
frequencies there is still some structure present in the P$_{3/2}$ channel,
which is the channel receiving contributions from all first peak
channels. See Fig.~\ref{sp3_tau2}, where we plot ATI spectra for the P
channels for different pulse lengths. As for the two--photon peaks,
also here the structures are washed out for shorter pulses.
In general the ATI peaks are at frequency $\omega_1+3\Omega$ and have
a bandwidth larger than that of the $\omega_1+2\Omega$ peaks. Infact they
entail the Fourier transform of the field pulse to the third power,
which has a smaller width (in time) than the square of the pulse.

   We note that the broadening of the ATI peaks was reported in the experimental
results reported in 1992 by Nicklich {\it et al} in
\cite{nick-etal:92}.  
The calculation described by these authors did not predict broadening of the higher peaks, however, although their spectra did 
contain the oscillations similar to those in the experiment, while those
oscillations are absent from our work.

\begin{figure}
\centerline{\psfig{figure=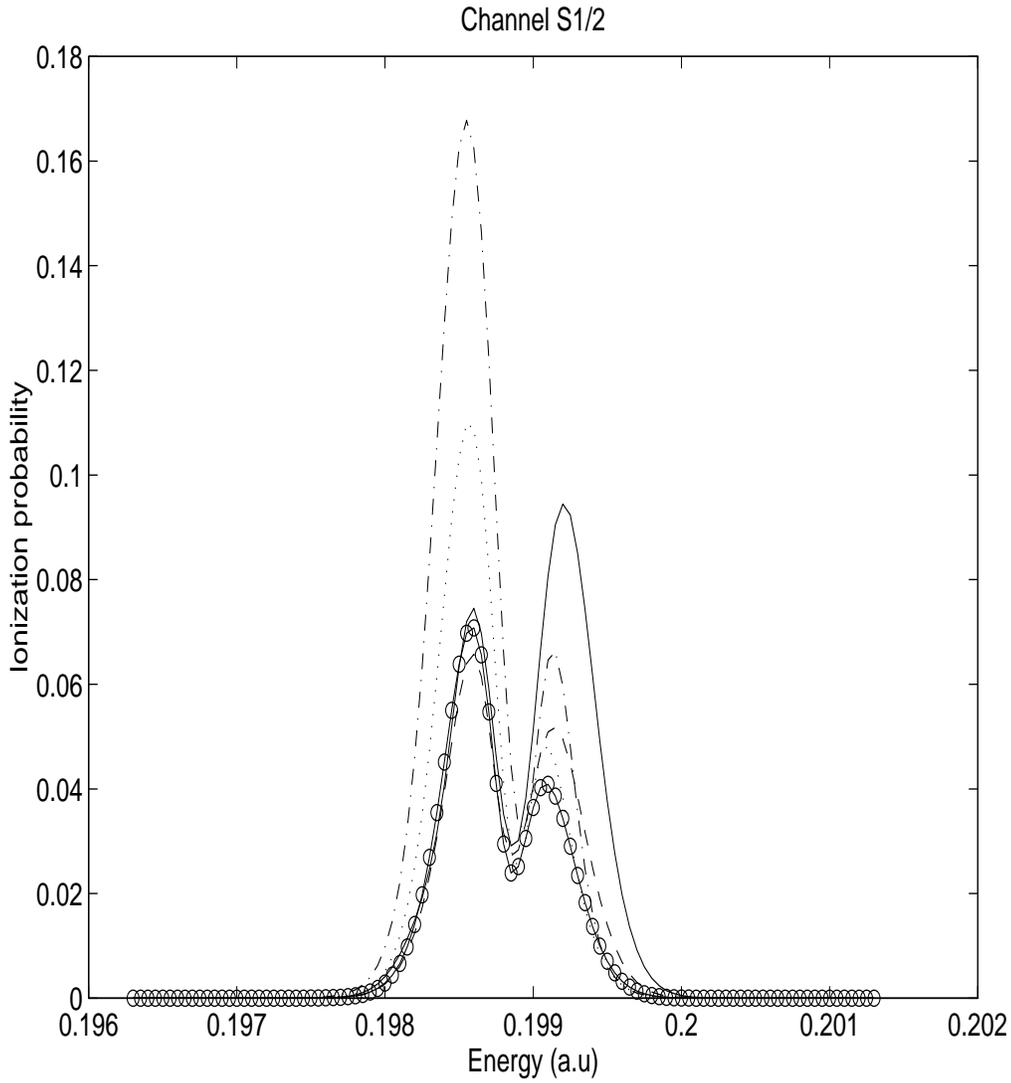,height=5.7in,width=5.7in}}
\caption{Probability per unit energy vs. energy for the photoelectron
produced by a two photon ionization of cesium for different chirping
parameters. The final continuum state is an S$_{1/2}$ channel.
Laser frequency 0.0994
a.u., pulse length 250 fs, peak intensity $10^{-7}$ a.u. ($1.4\times 10^{10}\rm{W/cm}^2$),
line type correspond to different chirp parameters $\beta$:
$-$ -0.0003 a.u., $--$ -0.0002 a.u., o  0.0 a.u.,
$\cdot\cdot$ 0.0002 a.u., $-\cdot$ 0.0003 a.u.}
\label{sp1_chirp2}
\end{figure}

The double peak structure is heavily influenced by the presence of chirping
in the laser light, especially for the longer pulses in the above range.
Varying the strength and the sign of the chirping changes the absolute and 
relative height of the peaks, see Fig. \ref{sp1_chirp2}.
In fact, the height of the peaks is determined by the amplitudes of the
excited states and their phase relationship, which are influenced by a kind of self--induced phase
modulation. Thus imposing external phase modulation (through chirping) on
these can strongly influence the ionization probability.

The evolution of the population of the bound states also changes appreciably
with chirping for the 7p levels which are quasi--resonant with the laser radiation.
The 6p levels are only slightly affected (see Fig. \ref{popP_chirp2}).
Infact the 6p's are off--resonance and their populations follow the
field adiabatically.
In addition the relative height of the side peaks is related to the
amount of population in the corresponding level.
The sign of the chirping determines which levels will have a bigger
average population during the pulse and thus the height of the
spectral peak. With our conventions a positive chirp corresponds to a
pulse whose instantaneous frequency is lower than the carrier
frequency at the beginning of the pulse and increases to a value
greater than the carrier frequency in the second half of the pulse.
Thus the 7p$_{1/2}$ is closer to resonance at the beginning of the
pulse and starts becoming populated earlier in the pulse. The
7p$_{3/2}$ on the contrary becomes populated at a higher rate later
during the pulse, when it gets closer to resonance. Thus its average
population during the pulse is lower than in the case of chirp free
light.
The situation is reversed for negative chirping.

\subsection{Strong Field Limit}

\begin{figure}
\centerline{\psfig{figure=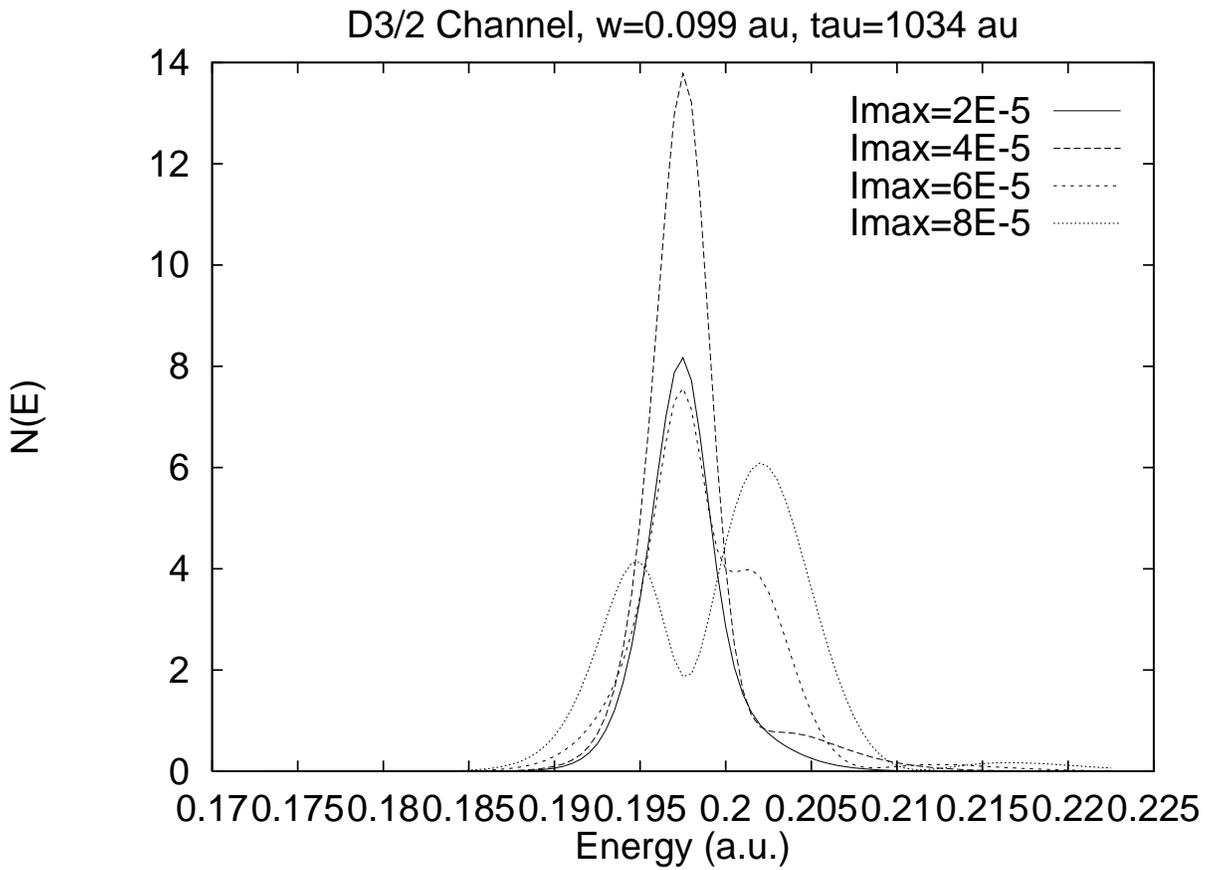,height=4.5in}}
\caption{Probability per unit energy vs. energy for the photoelectron
produced by a two photon ionization of cesium for various intensities.
The final continuum state is an D$_{3/2}$ channel.
The laser is tuned outside the 7p doublet, at the frequency 0.099
a.u., the pulse length is 25 fs. The peak intensity in the legend is
given in atomic units.}
\label{int1_D32}
\end{figure}

\begin{figure}
\centerline{\psfig{figure=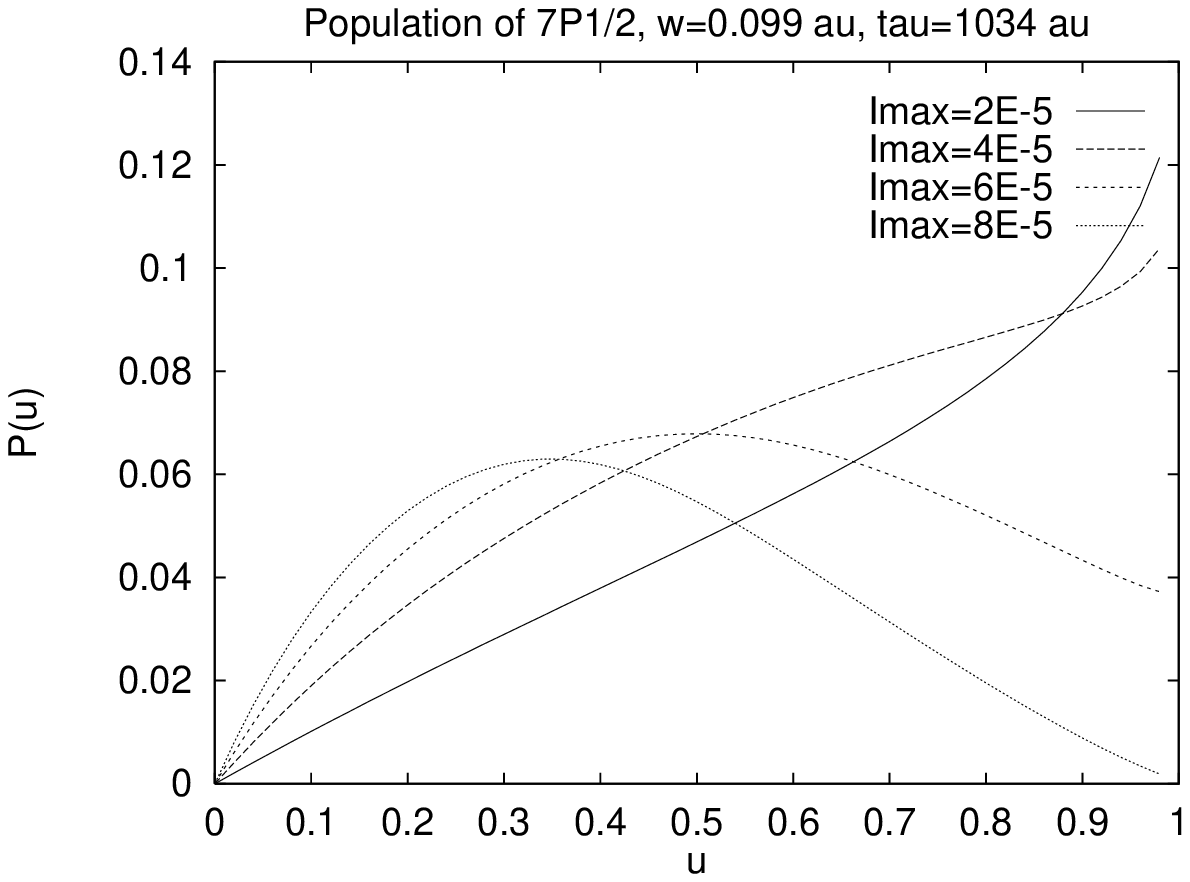,height=3.0in}}
\centerline{\psfig{figure=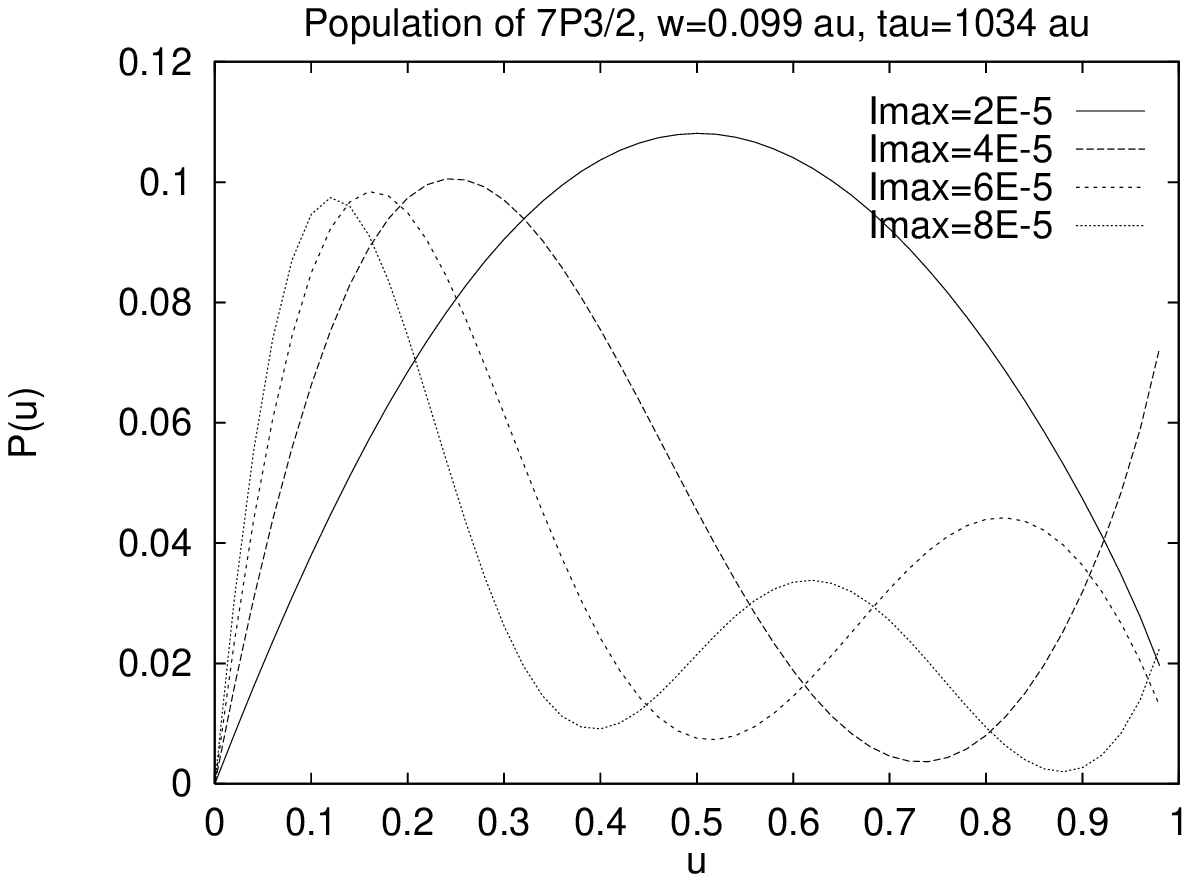,height=3.0in}}
\caption{ Population of the 7p levels vs compressed time, for various
intensities. The parameters are the same as in fig. \ref{int1_D32}
}
\label{int1_p}
\end{figure}

In the previous section we observed that the double peak structure
disappears for short pulses. However, Adler {\it et al.} \cite{adle-etal:95b} have
shown that when the pulse is short and 
the intensity of the beam increased, the spectra again contain a multipeak
structure, which is also present when the laser frequency is tuned
outside the doublet. Multiple peaks can appear also in the D$_{5/2}$ channel.
We can see this in Fig.~\ref{int1_D32}, where we
plot the channel D$_{3/2}$ spectra for different
intensities. Increasing the laser intensity we see a multipeak
structure emerge, at the same time the populations also start showing 
oscillations, see Fig.~\ref{int1_p}. 

As noted in the previous subsection,
this result confirms that
other mechanisms such as Rabi oscillations and Stark shifts play a
role in the production of the multipeak structure in strong fields.

As in the case of weak fields, for such short pulses, the influence
of chirping is minimal in the second continuum peaks, but it induces
visible changes in the population dynamics 
and the relative peak heights of the doublet.  

\begin{figure}
\centerline{\psfig{figure=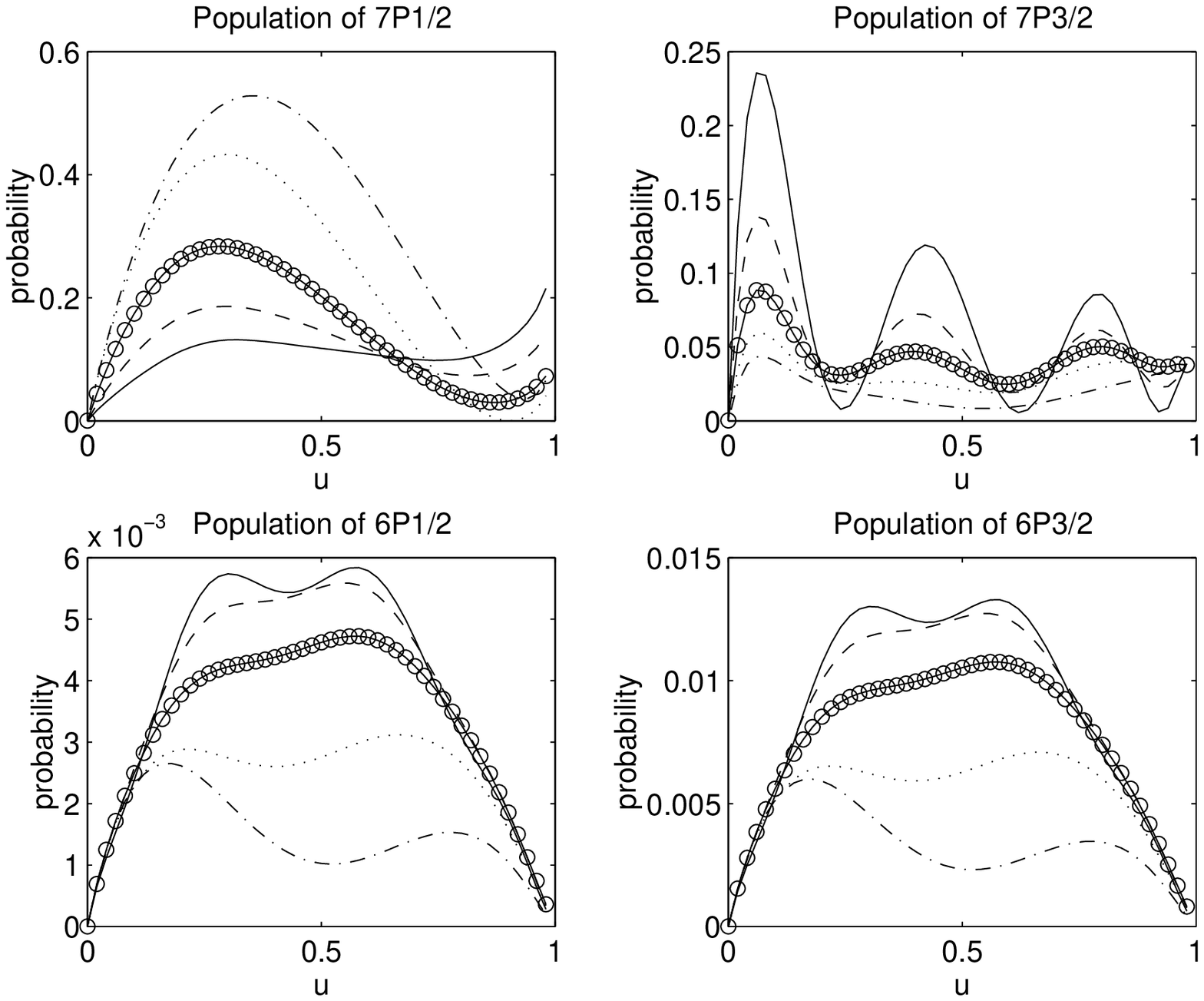,height=6in,width=5.7in}}
\caption{Populations of the 7p levels vs compressed time for
different chirp parameters.
Laser frequency 0.0994
a.u., pulse length 250 fs, peak intensity $0.7\times10^{-5}$a.u.,
line type correspond to different chirp parameters $\beta$:
$-$ -0.0003 a.u., $--$ -0.0002 a.u., o  0.0 a.u.,
$\cdot\cdot$ 0.0002 a.u., $-\cdot$ 0.0003 a.u.}
\label{fig:poP_c1}
\end{figure}

\begin{figure}
\centerline{\psfig{figure=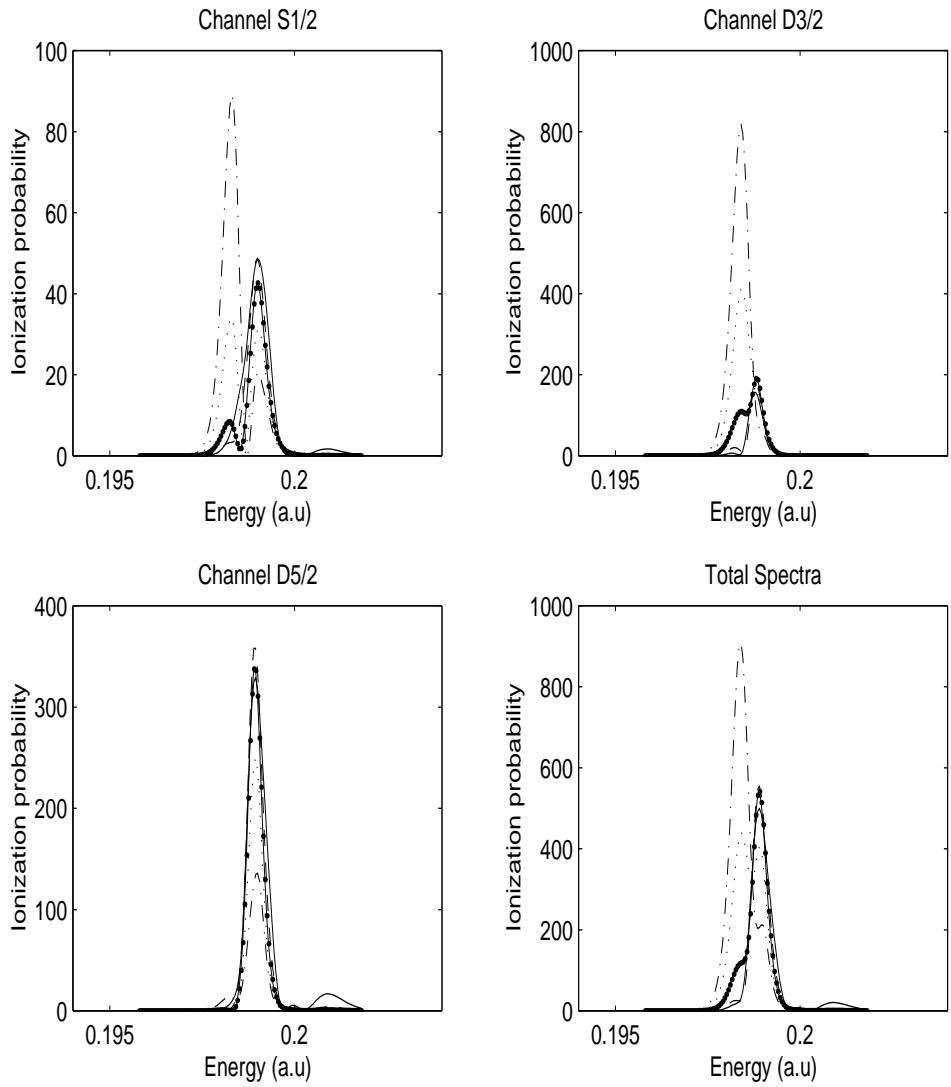,height=5.7in,width=5.7in}}
\caption{Probability per unit energy vs. energy for the photoelectron
produced by a two photon ionization of cesium for different chirping
parameters. The final continuum states are the S$_{1/2}$, D$_{3/2}$
and D$_{5/2}$ channels and the total probability for 2 photon ionization.
Laser frequency 0.0994
a.u., pulse length 250 fs, peak intensity $0.7\times10^{-5}$ a.u.,
line type correspond to different chirp parameters $\beta$:
$-$ -0.0003 a.u., $--$ -0.0002 a.u., solid/dotted line  0.0 a.u.,
$\cdot\cdot$ 0.0002 a.u., $-\cdot$ 0.0003 a.u.}
\label{fig:sp1_c1}
\end{figure}

\begin{figure}
\centerline{\psfig{figure=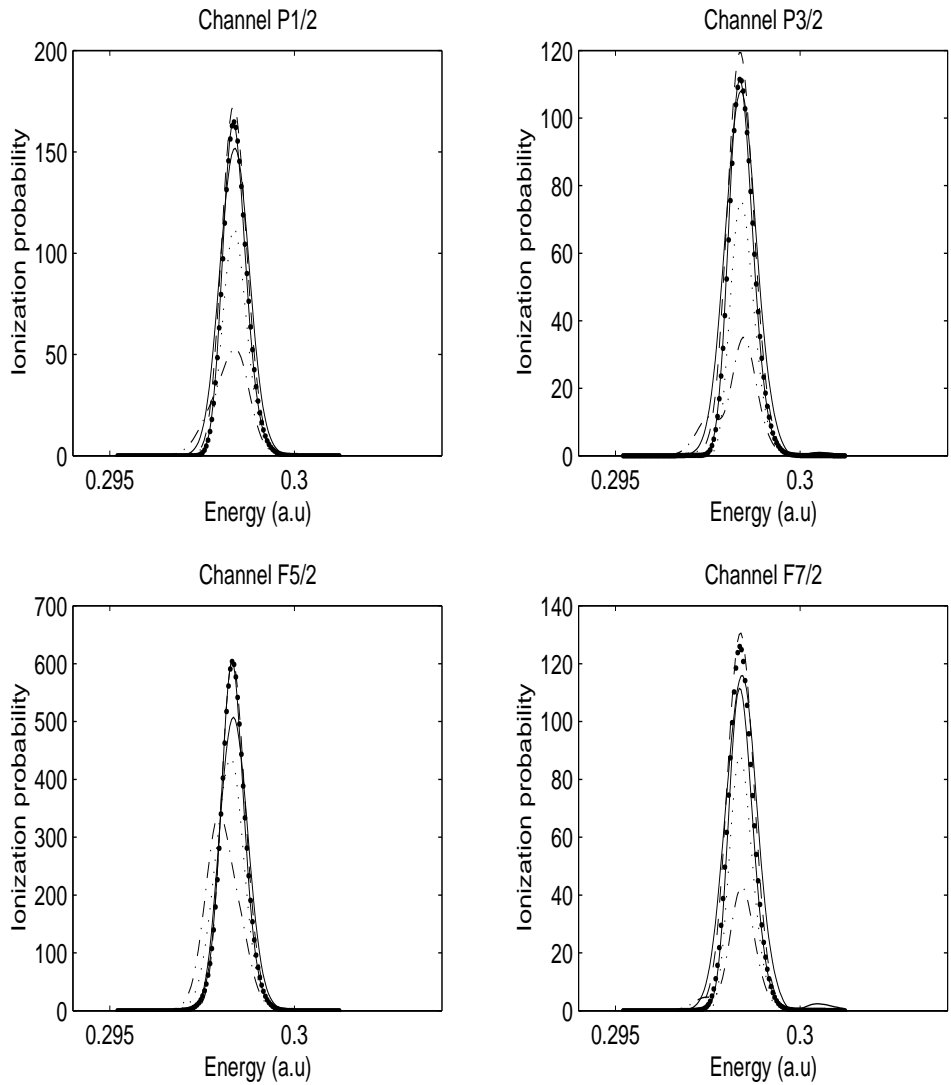,height=5.7in,width=5.7in}}
\caption{Probability per unit energy vs. energy for the photoelectron
produced by a three photon ionization of cesium for different chirping
parameters. Laser frequency 0.0994 a.u., pulse length 250 fs, 
peak intensity $0.7\times10^{-5}$ a.u.,
line type correspond to different chirp paprameters $\beta$:
$-$ -0.0003 a.u., $--$ -0.0002 a.u., solid/dotted line 0 a.u.,
$\cdot\cdot$ 0.0002 a.u., $-\cdot$ 0.0003 a.u.}
\label{fig:sp2_c1}
\end{figure}

For long pulses, the chirping causes dramatic changes in the
population dynamics and in the spectra of the threshold peaks, see Figs. \ref{fig:poP_c1},
\ref{fig:sp1_c1}. The ATI peaks are instead not strongly affected as
one expects because of the larger linewidth and general lack of
structure, as shown in Fig.~\ref{fig:sp2_c1}.
It appears that chirping enhances or quenches the oscillations in the
populations depending if it is an up--chirp or a down--chirp, while
the frequency of the oscillation remains the same.

\subsection{Gaussian Beams}
\label{ssec:4.3}

In actual experiments the laser beam incident upon the atomic sample
has an intensity spatial profile, so that the peak field seen by each
bound system is a function of its distance from the axis.
In weak fields this is not important, since the intensity is merely an
overall factor. 
This is not the case in strong fields, however.
Both the total ionization and the electron energy distribution become
complicated functions of the optical field strength, as for
example, in Fig.~\ref{int1_D32}
in which we see a second peak
developing in the D$_{3/2}$ channel with increasing intensity. 

If the intensity of the field is weak enough to allow the neglect of
depletion of the ground state and coupling between the excited states
through multiple transition to the  ground state and back, then none
excited states is affected by the existence of the other ones. 
In this limit the solution of each bound state amplitude takes the form
of the Rosen--Zener solution (see for example~\cite{robiscoe:78}
and~\cite{rose-zene:32}) for a two--level system in which all
energy shifts have been neglected.
The spectra produced by this expression are identical to those
produced by the exact result for weak laser intensities.

We examine for what intensities and pulse lengths
perturbation theory breaks down.
The evolution of the bound states start to deviate
from the typical behavior at weak fields when the peak intensity and
pulse length are such that $I_{\max} \tau \approx 10^{-3}$ a.u., 
for which nonlinear effects start to show.
This onset of nonlinear effects seems to be approximately independent of chirping.

In the perturbative regime, the total ionization is
proportional to the square of the intensity, since the
dominant process is two--photon absorption.

\begin{figure}
\centerline{\psfig{figure=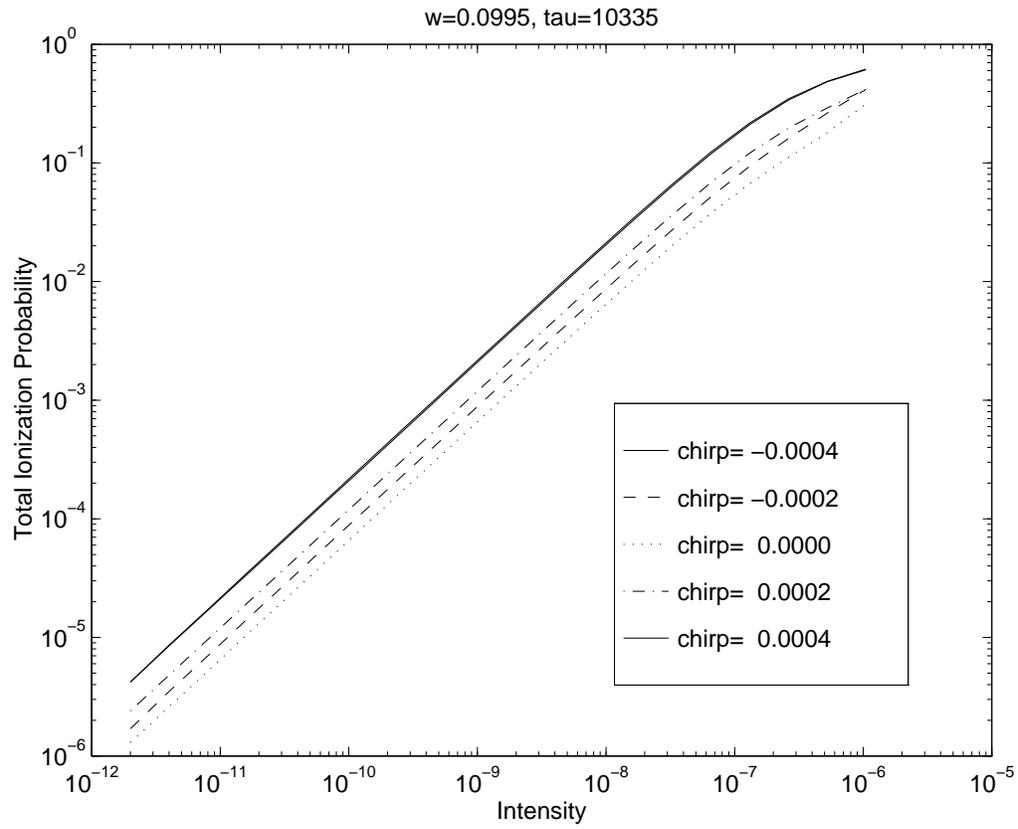,height=4.5in}}
\caption{Total ionization probability as a function of intensity for
different chirpings (in a.u.). The laser frequency is 0.0995 a.u. and
the pulse length 250fs.}
\label{fig:tion_t2}
\end{figure}

The total ionization probability is not very sensitive to the
presence of chirping, (Fig. \ref{fig:tion_t2}). There is in general a small 
increase in the ionization rate, which can be attributed to the widening 
of the pulse bandwidth from the frequency chirp.
The scaling parameter is $2 I_{\rm max}\tau/\pi$ and the total
ionization is practically the same when the length of the pulse or the
current $I$ is
changed maintaining the scaling parameter constant.

In the figures we show the evolutions of the populations
and the spectra as a function of the distance from the
center of the laser beam.

\begin{figure}
\centerline{\psfig{figure=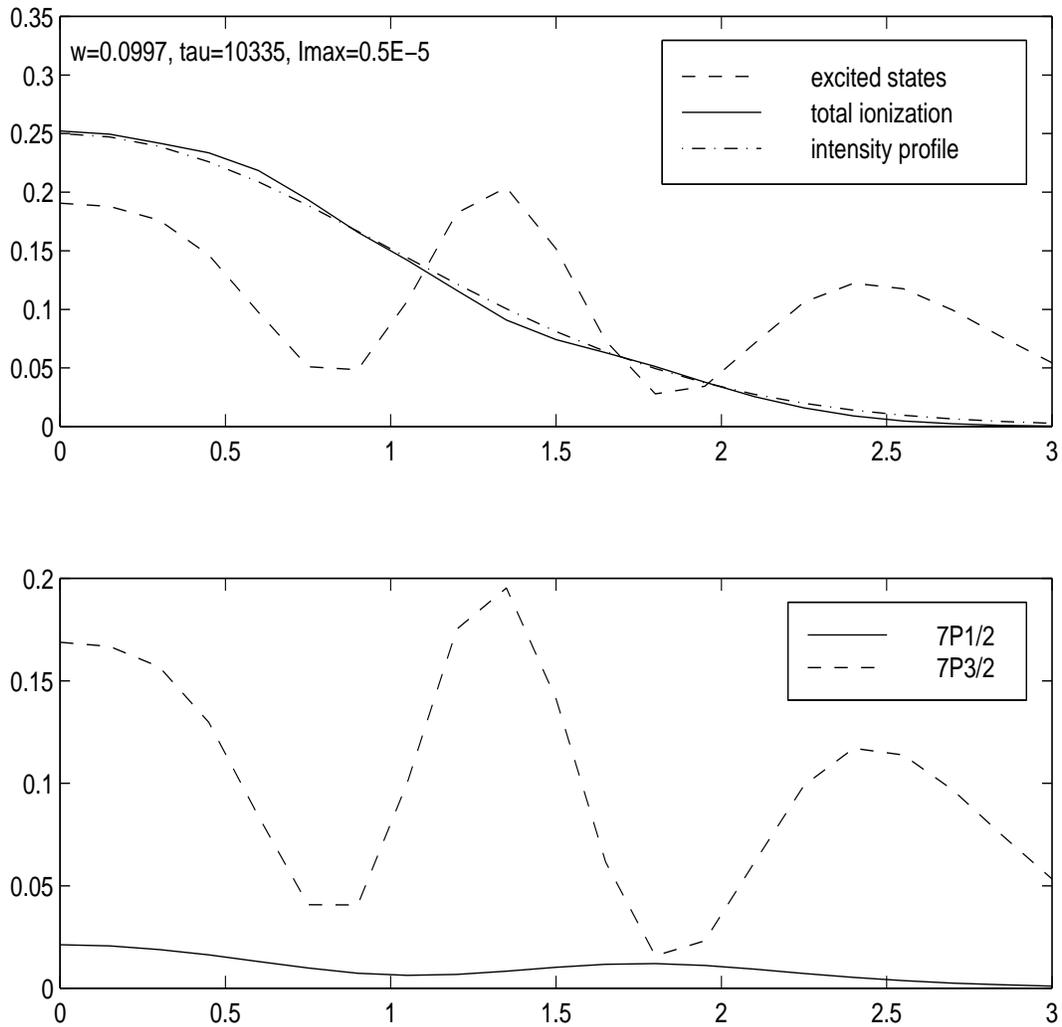,height=6in,width=5.7in}}
\caption{Residual total population of the excited states and total
ionization probability vs. distance r from the axis
of a Gaussian laser beam. The Gaussian beam profile is shown for
reference (dash--dotted line). Laser frequency 0.0997 a.u., pulse length
250 fs, peak intensity 0.7$\times 10^{12}$ W/cm$^2$, no chirping.}
\label{fig:res4}
\end{figure}

\begin{figure}
\centerline{\psfig{figure=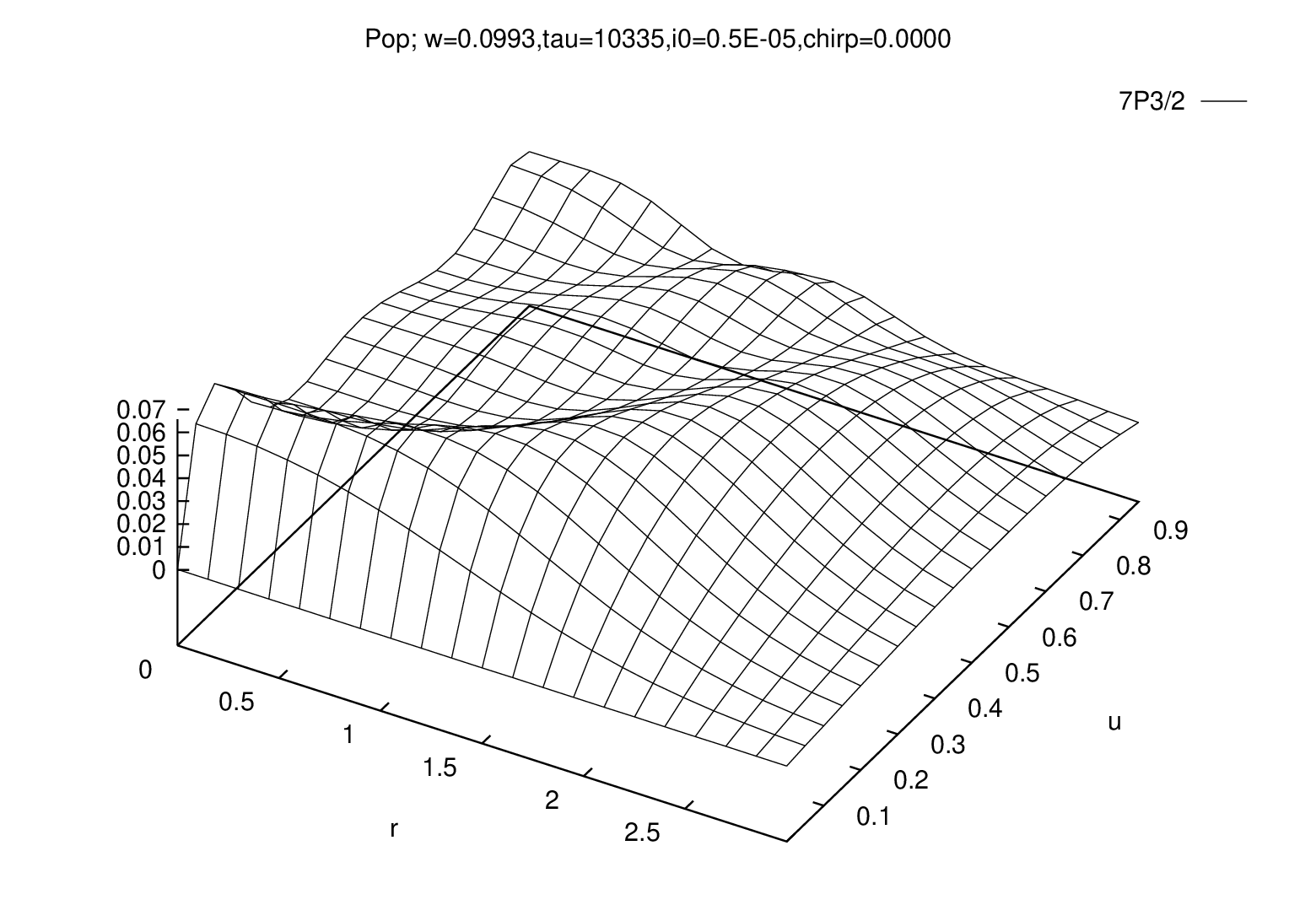,height=6in,width=5.7in}}
\caption{Population vs. compressed time u and distance r from the axis
of a Gaussian laser beam. Laser frequency 0.0993 a.u., pulse length
250 fs, peak intensity 0.7$\times 10^{12}$ W/cm$^2$, no chirping.}
\label{fig:po3_3d}
\end{figure}

\begin{figure}
\centerline{\psfig{figure=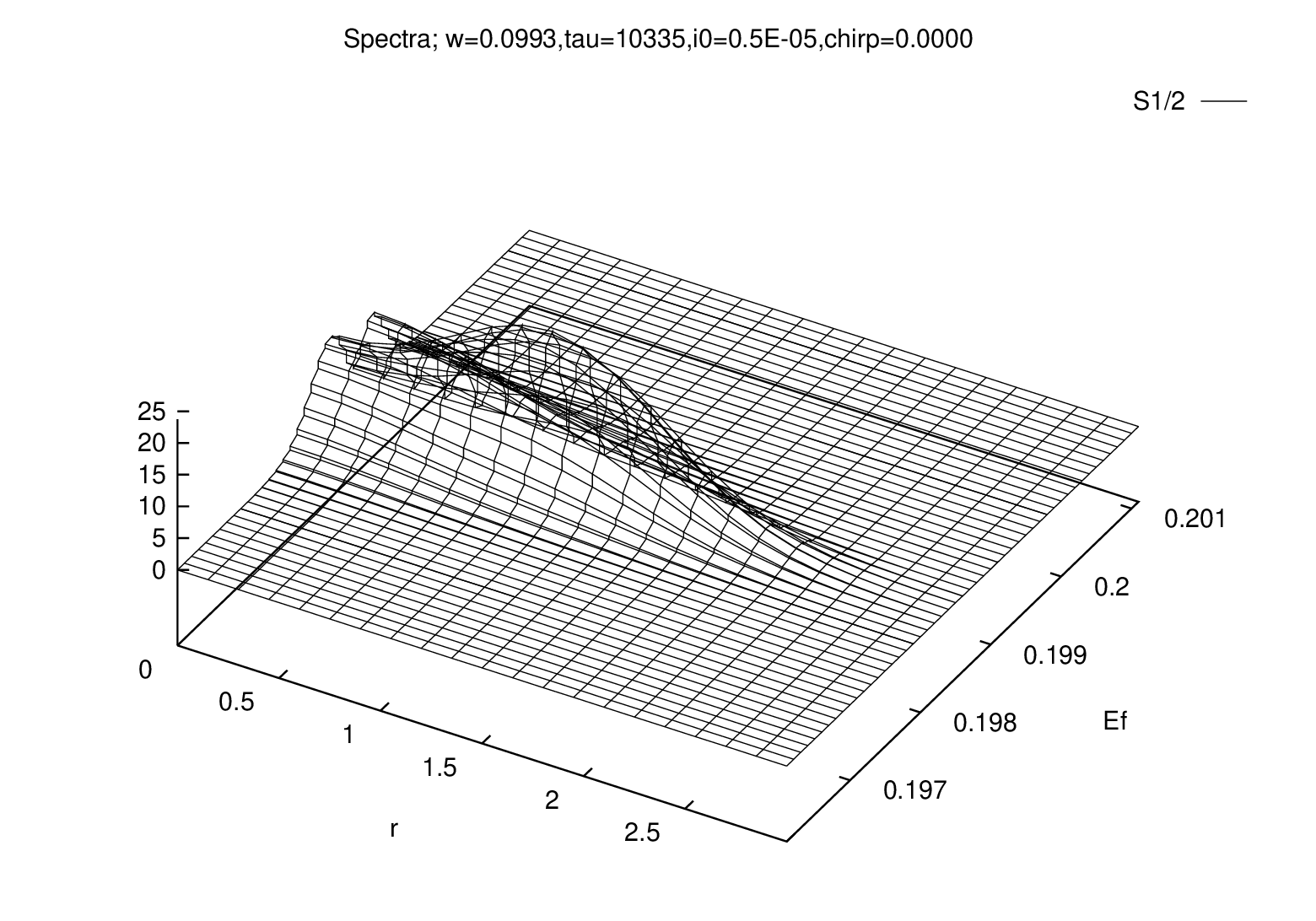,height=5.7in,width=5.7in}}
\caption{Probability per unit energy vs. energy for the photoelectron
produced by a two photon ionization of cesium vs. energy Ef and
distance r from the axis of a Gaussian laser beam.
 Laser frequency 0.0993 a.u., pulse length
250 fs, peak intensity 0.7$\times 10^{12}$ W/cm$^2$, no chirping.}
\label{fig:sp1_3d}
\end{figure}

\begin{figure}
\centerline{\psfig{figure=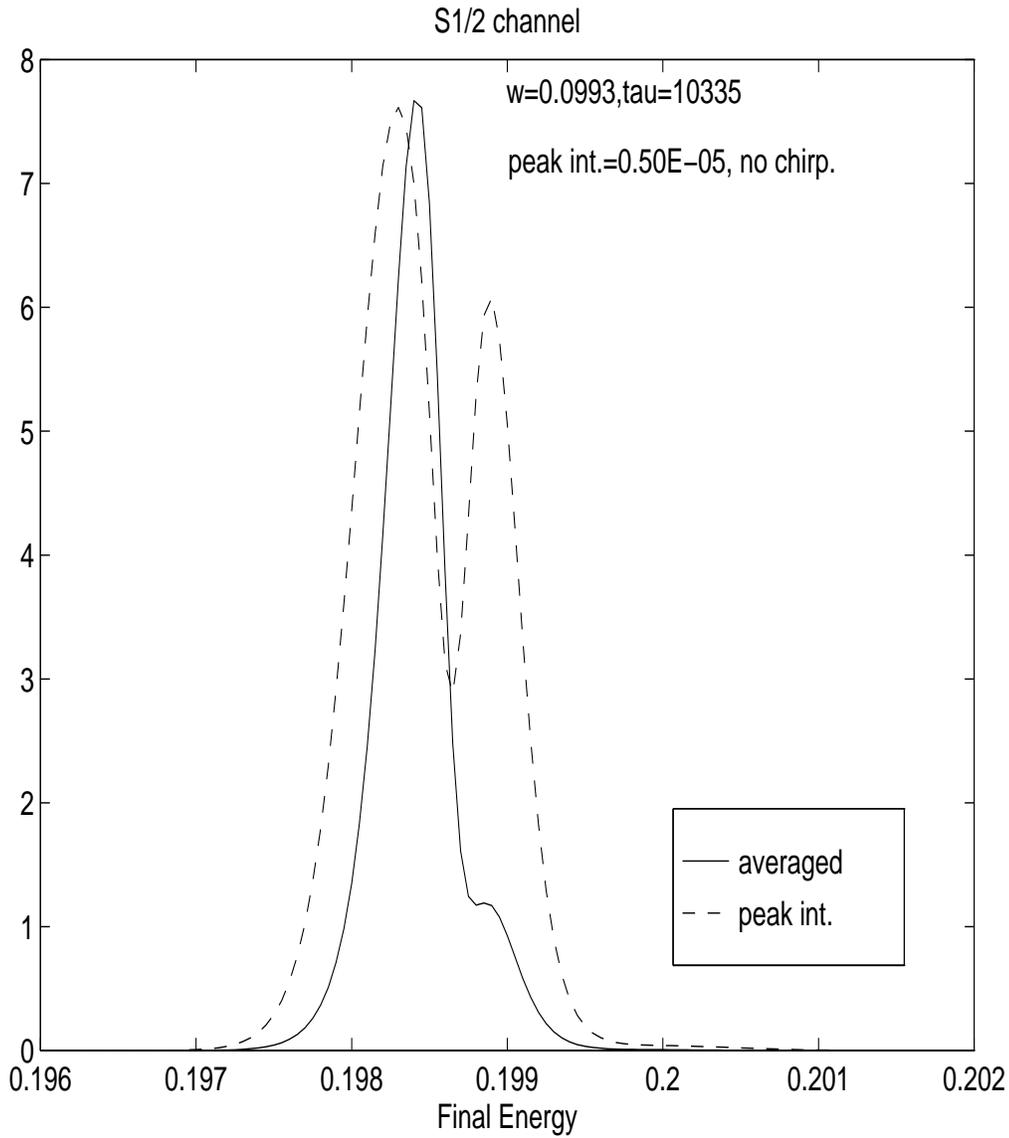,height=6in,width=5.7in}}
\caption{ Probability per unit energy for the photoelectron
produced by a two photon ionization of cesium vs. energy Ef.
The final continuum states are the S$_{1/2}$
 Laser frequency  0.0993 a.u., peak
intensity 0.7 $\times 10^{12}$ W/cm$^2$, pulse length 250 fs, no chirping.
Space averaged spectra (continuous line) and peak
intensity spectra (dashed line). }
\label{fig:avg_peak}
\end{figure}

Figs.~\ref{fig:res4} 
 show the total ionization probability and the
excited state population after the passage of the pulse, as a
function of the distance from the center of the beam. A curve showing
the Gaussian shape of the beam spatial profile is shown for reference.
We find that in general the total ionization probability roughly
follows the intensity profile of the laser beam, while the population
of the excited states after the passage of the pulse shows an
oscillatory dependence on the laser peak intensity at the location of
the atom. 
Looking at the temporal evolution of the excited states populations
one can see that the gradual change is due to the change in the Rabi
frequency and the number of oscillations the populations can do during
the pulse.
As an illustration in Figs.~\ref{fig:po3_3d} and 
we show the population of the 7p$_{3/2}$
state as a function of the distance from the center of the laser
beam without chirping and with chirping.
The distance is normalized such that the FWHM of the beam is unity with
the laser parameters given in the figure caption. 
The effect of chirping is
mainly to change the height of the oscillations, a chirping of the
opposite sign would flatten out the surface of the plot.
In Fig.~\ref{fig:sp1_3d} and 
we show the spectra for the s$_{1/2}$ channel for the
same set of parameters as Figs.~\ref{fig:po3_3d} 
respectively. The presence of chirping here changes the shape and
heights of the peaks.
One can see that for this intensity and frequency the
spectra does not depend in a monotonic way on the intensity.
The final spectra will be a spatial average of all these contributions
and the final population of the bound states will also
change with position.
Then in Fig. \ref{fig:avg_peak} we show a comparison between the spatially averaged
spectra and the spectra at the peak intensity.

\section{Conclusion}
\label{sec:5}
   The goal of this work was to model the dynamics of an atom with a finite
number of active levels interacting with short pulse of strong laser light
using a non-perturbative approach in the dominant states.  
It was hoped to strike a balance between
the competing needs of realism and numerical simplicity, extending prior
analysis to the case of ATI and a ``chirped'' pulse.

    As was done in prior work, we assumed only a few active levels whose
coupling to other states was taken into account by means of time--dependent
level shifts and decay rates.  By assuming a hyperbolic secant pulse shape
for the amplitude of the field, and a related form for the frequency modulation,
we were able to extend the analytically solvable problem previously reported
to the present case.  The ATI was accounted for by means of an effective
operator, whose coupling strength was obtained from the solution of an in-
homogeneous Schr\"odinger equation.

The model was applied to the cesium atom for light nearly resonant with
transitions between the 6s ground state and 7p excited doublet.   Coupling to
the 6p doublet is also included explicitly, since, even though its frequency
is far from resonance, the interaction matrix element is very large.  The
approach could be further extended to include more atomic levels and a larger
number of ATI peaks.
Allowance for the temporal profile of the laser beam was made, and the dynamics
of the excited state populations and spectra of the ejected electrons were
efficiently calculated.

   We found that chirping as a great influence on the dynamics of the quasi-
resonant states, but much less of an effect on the off resonant states.  This
transpires because the chirping causes much less of a fractional change in the
detuning of the nonresonant states.  In fact, it appears that the dominant
part of the change produced in the nonresonant amplitudes by the chirping is
due to the modification in the ground state amplitude cause by the resonant
intermediate couplings.

   The effect of chirping is more pronounced for longer pulses.  In this case,
the pulse has a smaller linewidth the the fractional change due to the chirping
is relatively more prominent.

  Some of the spectral features, such as double peaks, can change drastically
with the chirpings.  Those effects in weak fields are due to the interference
between the phases of the various atomic amplitudes when the laser is tuned
between resonances, and is very sensitive to time dependent frequency shifrts.
The interference is enhanced or suppressed depending on the sign and magnitude
of the chirp.

   The structure of the spectral peaks is quite sensitive to the length of the
pulse.  The doublet structure disappears for short pulses.  The shorter pulses
have a broader frequency spectrum, over which the populations dynamics are
averaged, thus washing out the deleicate interference effect due to phase
modulation.

   It is interesting to note that the ATI peaks lack the structure of the
principal two--photon peaks, even though both types are generated through the
same quasi--resonant intermediate states. 
This appears to result from the following considerations.  
The weak--field structure in the the two photon case stems from the
interference between the p1/2 and p3/2 amplitudes in a case when they are
tending to cancel, so that the overall transition probability is small.  In
the three--photon peak, there are more channels, so that the net probability
is dominated by the contributions that are not supressed by destructive in-
terference.

   However, according to our calculations, the three-photon peak is broader
than the two photon by about a factor two.  This can be understood in terms
of the Fourier transforms of the effective operator coupling the intermediate
states to the respective continua.  For two photon ionization, that operator
is just the transform of the hyperbolic secant, while for the ATI peak, it
is the transform of the square of the hyperbolic secant, which is broader in
frequency space.

   We note that chirping affects not only the shape of the photoelectron
structure , but also the total ionization probability, perhaps by making more
frquencies available for ionization.

We also studied the effect of spatial dependence of
the laser beam profile on the quantities accessible by experiment.
The total ionization, the spectra and the residual population show
deviations from perturbative behaviour which should be taken into
account performing a spatial average, to find the measured quantity.
The effects of a gaussian beam profile was discussed in
\cite{edwa-clar:96} for the case of a one--dimensional atomic
model. There the concurrence of significant amounts of ionization and
residual excited--state populations is attributed to the fact that
atoms are subjected to different peak intensities depending on
position, also oscillations in the residual excited populations are
found as a function of intensity.

As discussed previously we also find that for a certain range of intensities
and pulse lengths, the total ionization is proportional to the product
of peak intensity and pulse length. This seems to imply that the total
ionization is behaving like a one--photon absorption out of the p
states, and that this step is roughly given by first order
perturbation theory.

Assuming a Gaussian beam, the space averaged spectra can show
qualitatively  
 different features from the spectra corresponding to the peak
intensity.
In fact some of the spectral features are very sensitive to the peak
intensity. 
The residual excited state population shows an oscillatory
dependence on the laser intensity, while the total ionization
increases monotonically with the intensity.

In summary, we have shown that the methods used by Adler {\em et al.}
can be extended to the case of chirped pulses and to the inclusion of
ATI.
The population dynamics and ionization
spectra of cesium atoms subjected to strong short laser pulses has
been analyzed.
We have found that the behavior of that system is quite sensitive to
the laser light characteristics, such as
pulse lengths, frequency chirp and intensity spatial profile,
that are not usually taken into account.

\ack 

The authors wish to thank Marina Mazzoni and Guido Toci for helpful
discussion on the frequency chirping in short laser pulses.

\Figures

\clearpage

\end{document}